%% file: main.tex
\documentclass[sigconf]{acmart}
\input{macros}

\AtBeginDocument{%
  }

\setcopyright{acmlicensed}
\copyrightyear{2018}
\acmYear{2018}
\acmDOI{XXXXXXX.XXXXXXX}
\acmConference[Conference acronym 'XX]{Make sure to enter the correct
  conference title from your rights confirmation email}{June 03--05,
  2018}{Woodstock, NY}
\acmISBN{978-1-4503-XXXX-X/2018/06}

\begin{document}
\title{Retrieval-Augmented Test Generation: How Far Are We?}

\author{Jiho Shin}
\email{jihoshin@yorku.ca}
\orcid{0000-0001-8829-3773}
\affiliation{%
  \institution{York University}
  \city{Toronto}
  \state{ON}
  \country{Canada}
}

\author{Nima Shiri Harzevili}
\email{nshiri@yorku.ca}
\affiliation{%
  \institution{York University}
  \city{Toronto}
  \state{ON}
  \country{Canada}
}
\author{Reem Aleithan}
\email{reem1100@yorku.ca}
\affiliation{%
  \institution{York University}
  \city{Toronto}
  \state{ON}
  \country{Canada}
}

\author{Hadi Hemmati}
\email{hemmati@yorku.ca}
\affiliation{%
  \institution{York University}
  \city{Toronto}
  \state{ON}
  \country{Canada}
}
\author{Song Wang}
\email{wangsong@yorku.ca}
\affiliation{%
  \institution{York University}
  \city{Toronto}
  \state{ON}
  \country{Canada}
}

\renewcommand{\shortauthors}{Trovato et al.}

\begin{abstract}
\input{sections/0.abstract}
\end{abstract}
\RenewDocumentCommand{\respto}{m}{} 
\setcopyright{none} 
\settopmatter{printacmref=false} 
\renewcommand\footnotetextcopyrightpermission[1]{}
\maketitle

\input{sections/1.intro}
\input{sections/2.background}
\input{sections/3.methodology}
\input{sections/4.results}
\input{sections/5.threats}
\input{sections/6.conclusion}

\bibliographystyle{ACM-Reference-Format}
\bibliography{main}


\end{document}

%% file: macros.tex
\usepackage{multirow}
\usepackage{soul}
\usepackage{hyperref}
\usepackage{fancybox}
\usepackage{enumitem}
\usepackage{graphicx}
\usepackage{color}
\usepackage{xcolor}
\usepackage{listings}
\usepackage{array}
\usepackage{amsmath}
\usepackage{fancyhdr}
\usepackage{xspace}
\usepackage{url}
\usepackage{tikz}
\usepackage{soul}
\sethlcolor{yellow} 
\usepackage{caption}
\usepackage{pgfplots}
\usepackage{balance}
\usepackage{subcaption}
\pgfplotsset{compat=1.18}
\usepgfplotslibrary{groupplots}
\usepackage{pgf-pie}
\usepackage{algorithm}
\usepackage{algorithmic}
\usepackage{adjustbox}

\usepackage{makecell}
\usepackage{tabularx}
\usepackage{tcolorbox}
\definecolor{grey}{rgb}{0.9,0.9,0.9}
\makeatletter
\newcommand{\mybox}[1]{%
	\setbox0=\hbox{#1}%
	\setlength{\@tempdima}{\dimexpr\wd0+13pt}%
	\begin{tcolorbox}[boxrule=0.5pt, colback=grey, arc=2pt,
		left=2pt,right=2pt,top=2pt,bottom=2pt,boxsep=2pt]
		#1
	\end{tcolorbox}
}

\definecolor{songcolor}{RGB}{191,191,191}
\newcommand{\todoc}[2]{{\textcolor{#1}{#2}}}

\newcommand{\todoblue}[1]{\todoc{black}{#1}}

\newcommand{\add}[1]{\todoblue{#1}}

\newcommand{\respto}[1]{
\fcolorbox{black}{black!15}{
\label{response:#1}
\bf
  \scriptsize Review {#1}}~
}



%% file: sections/0.abstract.tex
Retrieval Augmented Generation (RAG) has advanced software engineering tasks but remains underexplored in unit test generation. To bridge this gap, we investigate the efficacy of RAG-based unit test generation for machine learning (ML/DL) APIs and analyze the impact of different knowledge sources on their effectiveness.  

We examine three domain-specific sources for RAG: (1) API documentation (official guidelines), (2) GitHub issues (developer-reported resolutions), and (3) StackOverflow Q\&As (community-driven solutions). Our study focuses on five widely used Python-based ML/DL libraries, TensorFlow, PyTorch, Scikit-learn, Google JAX, and XGBoost, targeting the most-used APIs.

We evaluate four state-of-the-art LLMs: \emph{LLMs—GPT-3.5-Turbo, GPT-4o, Mistral MoE 8x22B, and Llama 3.1 405B}, across three strategies: basic instruction prompting, Basic RAG, and API-level RAG. Quantitatively, we assess syntactical and dynamic correctness and line coverage. While RAG does not enhance correctness, RAG improves line coverage by 6.5\% on average. We found that GitHub issues result in the best improvement in line coverage by providing edge cases from various issues. We also found that these generated unit tests \respto{6-2} \add{can help detect new bugs. Specifically, 28 bugs were detected, 24 unique bugs were reported to developers, ten were confirmed, four were rejected, and ten are awaiting developers' confirmation.}

Our findings highlight RAG's potential in unit test generation for improving test coverage with well-targeted knowledge sources. Future work should focus on retrieval techniques that identify documents with unique program states to optimize RAG-based unit test generation further.

%% file: sections/1.intro.tex
\section{Introduction}
\label{sec:intro}

Retrieval Augmented Generation (RAG) has seen increasing use in various automated software engineering tasks, including code generation \cite{su2024arks,rani2024augmenting}, code search \cite{gotmare2023efficient}, commit message generation \cite{zhang2024rag}, code suggestions \cite{chen2024code}, code completion \cite{tan2024prompt, liu2024graphcoder}, assertion generation \cite{nashid2023retrieval, quddus2024enhanced}, and automated program repair \cite{wang2023rap,nashid2023retrieval}, showing significant improvements.
\respto{8-2} \add{RAG is an approach that enhances LLMs by fetching relevant external documents and using them to ground and improve their responses \cite{lewis2020retrieval}.}
\respto{5-7} \add{
A key factor behind RAG's effectiveness is that although LLMs are trained on massive corpora, they often cannot fully leverage the knowledge when tackling complex tasks. Augmenting them with targeted data surfaces and grounds insights to a more correct response \cite{gao2023retrieval}.}
Despite its notable advancements, the potential of RAG in unit test generation remains underexplored. 
To bridge this gap, we investigate the efficacy of RAG in test generation. 
As RAG can be built on different knowledge bases, we explore the impact of different document sources on test generation to provide insights into its practical benefits and limitations. 

Specifically, we consider the effect of three types of domain knowledge for the RAG used in this work: 1) API documentation, 2) GitHub issues, and 3) StackOverflow Q\&As. We include these sources as they could provide examples of API usage for creating tests. 
Each source offers knowledge from different perspectives: API documentation with official usage guidelines, GitHub issues offer resolution from the library developers, and StackOverflow Q\&As present community-driven solutions and best practices.

In our experiments, we focus on five well-known Python-based Machine and Deep learning (ML/DL) libraries: TensorFlow, PyTorch, Scikit-learn, Google JAX, and XGBoost. The domains were chosen for two key reasons. First, they play a critical role in modern software development by providing essential tools, frameworks, and abstractions for building, training, and deploying complex neural networks efficiently \cite{abadi2016tensorflow,daghighfarsoodeh2025deep}. \respto{5-8} \respto{7-5} \add{Second, unlike traditional software, ML/DL systems are inherently statistical and non-deterministic. Their APIs reveal expansive, often effectively infinite parameter spaces and exhibit complex object behaviours, making them fundamentally different from conventional, deterministic code. These differences make unit testing more challenging, yet many ML libraries remain under-tested. This study demonstrates how RAG-based LLMs can systematically generate robust, targeted tests to improve the reliability of such frameworks~\cite{wang2021automatic, shin2023mlcodegen, wan2024keeper, xia2023balancing}.}

We evaluate the effectiveness of four state-of-the-art LLMs, i.e., \emph{GPT-3.5-Turbo, GPT-4o, Mistral MoE, and Llama 3.1}. 
In total, we inspect the unit tests generated for 694 APIs across the five ML/DL libraries, with 19K GitHub issues and 22K StackOverflow Q\&As.

We start by quantitatively evaluating the generated tests based on their Abstract Syntax Tree (AST) parse rates, execution rates, and pass rates. Next, we calculate the line coverage on the API under test. Finally, we conduct a manual analysis on a subset of the generated tests to evaluate the bug detection rate and the impact of different RAG sources on the software under test in greater depth.

Our findings indicate that LLMs with basic instruction prompting generate syntactically correct test cases with an average parse rate of 93.89\%, an execution rate of 71.35\%, and a pass rate of 61.11\%. When evaluating RAG-based unit test case generation, we observe promising results compared to basic instruction prompting. Specifically, RAG unit test generation improves line coverage by an average of 6.5\%, with the highest improvement reaching an average of 8.1\% when using API-level GitHub issue RAG.

We also analyzed the bug detectability from the generated unit tests and found that it detected 28 bugs, 24 unique bugs were reported to developers, so far, ten were confirmed.
Finally, we find that RAG helps cover unique lines by providing unique states of programs that cannot be generated directly by LLMs.

\respto{7-1} \respto{7-2} \add{This work pioneers RAG-based LLMs for unit test generation in ML/DL libraries and systematically evaluates how three domain-specific knowledge sources, i.e., official documentation and community-based forums, can advance code coverage and bug detectability. We also propose a simple yet effective RAG approach, API-level RAG, that improves test adequacy.}

This paper makes the following contributions:
\begin{itemize}
    \item We are the first to explore the potential of RAG-based LLMs for unit test generation tasks on ML/DL libraries.
    \item We investigate the effectiveness of RAGs with different sources of domain knowledge for unit test generation, i.e., API documentation, GitHub issues, and StackOverflow Q\&As.
    \item We perform a comprehensive manual analysis of the generated tests that detect bugs and examine the advantages and limitations of various domain knowledge sources in RAG-based unit test generation.
    \item We created a benchmark dataset for RAG-based test generation. All code and data used in the experiment are shared as a replication package for future researchers\respto{8-7}\add{\footnote{\url{https://github.com/shinjh0849/api_guided_testgen}}}.
\end{itemize}

We organized the rest of this paper as follows.
Section~\ref{sec:background} introduces the background of our study.
Section~\ref{sec:methodology} details the data collection and implementation of the RAG-based test generation.
Section~\ref{sec:results} shows the results of our experiments. 
Section~\ref{sec:threats} discusses the possible threats in the study.
Section~\ref{sec:conclusion} concludes this paper. 

%% file: sections/2.background.tex
\section{Background and Related Works}
\label{sec:background}
\subsection{RAG Incorporation in ASE}
Retrieval-augmented generation is a prompting strategy that enhances the responses of LLMs by incorporating additional knowledge sources beyond the pre-trained knowledge base~\cite{baumann2024combining}. The query intended for the LLM is used to search and retrieve relevant documents from a database~\cite{baumann2024combining, chen2024benchmarking}. This process leverages external knowledge, such as documents from various sources, to produce more contextually relevant and accurate responses compared to traditional generation-only models~\cite{chen2024benchmarking}.

One advantage of RAG is that it eliminates the need for retraining, as the LLM can search the latest information to generate more reliable output through retrieval-based generation~\cite{zhang2024enhancing}. These techniques have been successful in numerous software-related tasks, including code generation \cite{su2024arks}, code search \cite{gotmare2023efficient}, commit message generation \cite{zhang2024rag}, code suggestions \cite{chen2024code}, assertion generation \cite{nashid2023retrieval}, and automated program repair \cite{wang2023rap,nashid2023retrieval}. However, the effectiveness of RAG relies heavily on the quality of the retrieved knowledge~\cite{lin2024novel}. Therefore, it is essential to assess the impact of various knowledge resources on the performance of RAG-based techniques.

\subsection{Unit Test Generation using LLMs}
Recently, numerous attempts have been made to generate tests using LLMs automatically \cite{shin2024assessing, derakhshanfar2022basic, shin2024domain, yuan2024evaluating, chen2024chatunitest}.
Yuan et al. \cite{yuan2024evaluating} found that while \emph{ChatGPT} generates unit tests with good coverage and readability, it struggles with correctness. To address this, they propose \emph{ChatTester}, which enhances test correctness through iterative refinement.
Chen et al. \cite{chen2024chatunitest} introduced \emph{ChatUniTest}, an LLM-based unit test generation framework with an adaptive focal context mechanism and a generation-validation-repair workflow, enhancing accuracy and outperforming existing tools.
Yang et al. \cite{yang2024evaluation} empirically evaluated open-source LLMs for unit test generation, comparing them to \emph{GPT-4} and \emph{Evosuite}. They emphasize prompt design, hallucination issues, and in-context learning, concluding that while promising, LLMs are still outperformed by traditional methods in coverage and defect detection.
Zhang et al. \cite{zhang2024llm} introduced Property-Based Retrieval Augmentation, implemented in \emph{APT}, to enhance LLM-based unit test generation using task-specific code context and the Given-When-Then (GWT) structure. This approach improves correctness, completeness, and maintainability through iterative refinement and structured property retrieval.
Others studied test generation to increase their code coverage using search-based algorithms, symbolic execution, and genetic algorithms \cite{wang2021automatic,leitner2007contract, fraser2011evosuite, tonella2004evolutionary}. 
In contrast to these studies, our study focuses on the impact of different knowledge sources for RAG-based unit test generation with a specific emphasis on ML/DL libraries.

Apart from previous studies that focus on Java projects, studies on Python projects, specifically in the ML/DL domain.
Yang et al. \cite{yang2023fuzzing} introduced \emph{$\nabla$Fuzz}, the first general-purpose API-level fuzzer for testing automatic differentiation in deep learning libraries. It detects inconsistencies in higher-order gradients using differential testing across execution modes, outperforming prior fuzzing techniques in bug detection and code coverage.
Zou et al. \cite{zou2023deep} proposed \emph{Ramos}, a novel hierarchical heuristic test generation approach for ML projects. Their study employed a hierarchical structure of machine learning models to generate more models by mutation techniques. 
Narayanan et al.~\cite{narayanan2023automatic} proposed \emph{MUTester}, which mines API usage patterns from StackOverflow to enhance search-based unit test generation for ML projects.

\respto{7-4a} \add{Note that previous work on LLM-driven unit test generation has largely focused on iterative prompting strategies via execution feedback \cite{ryan2024code, chen2024chatunitest} or limited RAG without systematically exploring how different domain-specific context or community knowledge can improve coverage and bug detection \cite{narayanan2023automatic, yang2024evaluation}.}
\respto{5-6a} \add{In contrast, our study is the first to systematically investigate and compare the effectiveness of three distinct external knowledge sources as RAG databases for unit test generation, bringing new insights into their comparative performance.}

%% file: sections/3.methodology.tex
\section{Methodology}
\label{sec:methodology}
This section details the methodology of our study, i.e., the research questions, the data collection, the retrieval-augmented test generation framework, baseline LLMs, and the evaluation metrics. 
Figure \ref{fig:overall_fig} depicts the overall flow of our study. 


\subsection{Research Questions}
To investigate the effectiveness of RAG-based unit test generation for ML/DL libraries, we devised the following research questions:

\vspace{4pt}
\noindent{\textbf{RQ1: How effective are LLMs in generating unit test cases?}}\\
\noindent \textbf{Motivation.}
This RQ investigates the ability of state-of-the-art LLMs using basic instruction prompting to generate unit test cases without document augmentation. 
This RQ serves as a baseline for subsequent research questions.

\input{figure/overall_fig}

\vspace{4pt}
\noindent{\textbf{RQ2: How effective is the Basic RAG approach in enhancing LLMs' unit test generation using domain knowledge from API documentation, GitHub issues, and Stack Overflow Q\&As?}}\\
\noindent \textbf{Motivation.}
This RQ examines the impact of the Basic RAG approach using documents from three external sources. We investigate improvements in test generation quality, focusing on syntactical and execution correctness and test adequacy (line coverage). 

\vspace{4pt}
\noindent{\textbf{RQ3: How effective is the API-level RAG approach compared to the Basic RAG in the LLM-based unit test case generation?}}\\
\noindent \textbf{Motivation.}
In this RQ, we designed an API-level RAG database to improve upon the Basic RAG approach. While Basic RAG may retrieve irrelevant documents related to other APIs, our API-level RAG aims to ensure that only documentation relevant to the target API under test is retrieved, thereby improving precision and relevance.


\vspace{4pt}
\noindent \textbf{RQ4: How effective are the unit test cases generated by RAG-based LLMs in bug detection?}\\
\noindent \textbf{Motivation.}
In this RQ, we manually investigate how many true bugs can be detected when executing the generated unit tests to evaluate their practicality.

\vspace{4pt}
\noindent{\textbf{RQ5: What are the underlying reasons when a particular source of knowledge improves the baseline LLMs?}}\\
\noindent \textbf{Motivation.}
In this RQ, we conduct a manual analysis of a subset of APIs to further investigate the root causes of how and why these RAG sources and approaches influence unit test generation. 

\subsection{Data Collection}
\label{sec:data_collection}

\subsubsection{Studied ML/DL libraries}
For our experiments, we focus on five major Python-based ML/DL libraries, i.e., \textbf{TensorFlow}\footnote{\url{https://www.tensorflow.org/}} (abbr. \textbf{tf}), \textbf{PyTorch}\footnote{\url{https://pytorch.org/}} (abbr \textbf{torch}), \textbf{Scikit-learn}\footnote{\url{https://scikit-learn.org/stable/index.html}} (abbr. \textbf{sklearn}), \textbf{Google JAX}\footnote{\url{https://github.com/google/jax}} (abbr. \textbf{jax}), and \textbf{XGBoost}\footnote{\url{https://github.com/dmlc/xgboost}} (abbr. \textbf{xgb}). They were selected due to their widespread popularity and adoption in both academia and industry, their large and active communities with extensive documentation, and their significant roles in the field of machine and deep learning. These factors make them ideal candidates for a study aimed at improving test case generation using RAG.

\input{table/data}

\subsubsection{Experiment Data Collection.}
The collected data include 1) the list of API names, 2) API documentation, 3) GitHub issues and the responses, and 4) the StackOverflow Q\&As. Initially, we gathered all API lists from their respective \respto{5-1} \add{official API documentation website}. For API documentation, we extract 1) the signature information with parameter details, 2) the natural language description detailing the functionality of the API and its parameters, and 3) example code demonstrating correct usage patterns, if available.

For the second RAG source, we mine closed issues from their GitHub repositories, specifically extracting the title, description content, and associated comments. For the final RAG source, we collect relevant StackOverflow questions using the tag feature and the associated accepted answers by the poster. Each API document, GitHub issue thread, and StackOverflow Q\&A pair is treated as a separate document chunk for the RAG database. \respto{5-1} \add{To control dataset size and inference costs within budget and ensure compatibility with various LLM context windows, we truncated all documents to a maximum of 5K tokens. This prevents a few exceptionally long texts (which often come from lengthy issue threads) from exceeding context limits or incurring unnecessary computational overhead. We found that the latter parts of these threads (e.g., replies or solutions) are generally less relevant for test generation, though they may be useful for tasks like program repair.}

Then, we apply a simple string-matching heuristic to filter out documents irrelevant to API usage. Specifically, we only include GitHub issues and StackOverflow Q\&A documents that mention any of the APIs for the RAG document (either in the title or the content). We link the mentioned APIs to each document and use this mapping information for the API-level RAG approach (Details in \ref{sec:rag_approach}). Any APIs not mentioned in either source (issues and Q\&As) are excluded from data collection. This results in a many-to-many relationship between issues and Q\&A documents with APIs, while API documents maintain a one-to-one relationship.

The version of the API documentation collected is tf: 2.16.1, torch: 2.6.0, sklearn: 1.5.0, jax: 0.4.28, and xgb: 2.1.0, which was the stable version at the time of the conducted study.
The time frame of the GitHub and StackOverflow document collection is from the beginning to Jun. 2024.
The distribution of the number of APIs after filtering and the number of associated documents is presented in Table \ref{tab:data}.
\respto{5-2} \respto{5-9} \add{To ensure a fair and comprehensive evaluation of RAG across different external knowledge sources, we only considered those APIs for which both GitHub issues and StackOverflow Q\&A each contain at least ten documents (to ensure sufficient context for retrieval), yielding a final set of 694 APIs in total.}

\subsection{Retrieval-Augmented Test Generation}
\label{sec:rag_approach}
The RAG approach searches relevant documents and augments the query to assist LLMs in generating more effective and correct unit test cases for ML/DL APIs.
We consider two settings for the RAG approach, i.e., the Basic RAG and the API-level RAG.
\respto{1-1} \respto{1-4} \add{The key difference between Basic RAG and API-level RAG lies in the structure and scope of their retrieval databases. Basic RAG retrieves the top-k globally relevant documents from a shared corpus, regardless of whether they mention the target API. In contrast, API-level RAG constructs a separate database for each API using keyword search, then applies a dense retriever within that API-specific set to ensure more focused and relevant retrieval. This hybrid sparse and dense strategy was shown to outperform either approach alone in previous studies \cite{arivazhagan2023hybrid, wang2024searching}.} Here, a “dense retriever” refers to a neural embedding retriever (for example, a bi-encoder) that scores passages by vector similarity, while the “sparse” retriever uses the initial keyword or BM25 filter that narrows the API-specific corpus before dense retrieval or re-ranking.

\subsubsection{RAG Database Creation:} 
The Basic RAG approach stores all documents in a single database, reflecting the most common cases of RAG-based LLMs~\cite{wang2024coderag, chen2024code, nashid2023retrieval}. 
\respto{1-1} \respto{1-4} \add{Basic RAG retrieves the top-k documents from a shared database that aggregates documents related to all APIs across the five ML/DL libraries for each source (API documentation, GitHub Issues, StackOverflow Q\&As, and their combination).}
To investigate the impact of different RAG sources, we construct four types of Basic RAG databases: 1) all API documents combined, 2) all GitHub issues combined, 3) all StackOverflow Q\&A combined, and 4) an aggregation of the three. As mentioned previously, we discard all documents irrelevant to API usage.  
When filtering these documents, we additionally associate documents with APIs using simple string-matching heuristics. We use this mapping information to construct the RAG database for each API, which strongly constrains the search to relevant documents. This method provides a solid constraint for retrieving only pertinent documents. Consequently, we create a database for each API to facilitate searching within when generating unit test cases for a particular API, thereby defining the API-level RAG approach. 

We construct the vector databases for the RAG using the  \emph{chromadb}\footnote{\url{https://www.trychroma.com/}} library, an open-source vector database designed for storing vector embeddings. It employs the nearest neighbour embedding search for efficient semantic search, utilizing Sentence Transformers~\cite{reimers2019sentencebert} to embed the documents, which is considered a dense retrieval method. We employ the dense retrieval method as prior work has generally shown it to be more accurate than sparse retrieval methods~\cite{parvez2021retrieval, sawarkar2024blended}. 

\input{figure/zeroshot_prompt}
\input{figure/augmented_prompt}

\subsubsection{Prompting LLMs:}
We task an LLM to generate unit test cases for each API to ensure maximum coverage of its functionality. Additionally, we require generating new tests only if they cover new lines, allowing the model to determine the number of test cases needed autonomously. The generated test suite must utilize the \emph{unittest} library and include all necessary imports and main functions to ensure the code is directly executable. 

\respto{2-1} \respto{2-2} \add{We followed a bottom-up, task-specific approach to design our prompts. First, we defined a system prompt framing the model as a test suite generator for a given ML/DL library, encouraging developer-style output. The instruction prompt was crafted to produce standalone/executable test suites, including all necessary imports and a main function. We also instructed the model to generate tests only if they cover untested lines. The prompt was refined through small pilot runs. We observed whether the outputs were executable, diverse, and free of redundancy, and iteratively tuned the phrasing until the results consistently met our criteria. Figure~\ref{lst:zeroshot_prompt} illustrates the prompt used for basic instruction prompting.} When augmenting the prompt with documents retrieved from the RAG database, we instruct the model to utilize the provided documents to enhance the test case's effectiveness and exercise more lines. The additional prompt we use for augmented generation is listed in Figure \ref{lst:augmented_prompt}. \respto{2-2} \add{A more detailed demonstration of the engineered prompts can be found in our replication package.}

For Basic RAG prompting, we set the number of retrieved documents to three, as previous research~\cite{chen2024code} indicated that metric improvements plateaued beyond this number. Specifically, we retrieve the top three API documents for Basic API-documentation RAG, the top three issue documents for Basic GitHub-issue RAG, and the top three Q\&A documents for basic StackOverflow Q\&A RAG. \respto{1-2} \add{For the Basic combined RAG, note that the top three documents retrieved are not necessarily from each of the three sources as the retrieval criterion is general relevance.}

To retrieve the documents, we use a query to search for relevant documents and specify the number of documents to retrieve. The query used for retrieval is the instruction prompt: ``Generate a Python unit test case to test the functionality of \emph{API\_NAME} in \emph{ML/DL\_LIB} library with maximum coverage.''

For API-level RAG prompting, we set different numbers of documents for retrieval: 1) one for the API document, as there is only one document per API, 2) three for GitHub issues and StackOverflow Q\&A, similar to Basic RAG prompting, and 3) one document each (total of three) for the combined API-level RAG approach. We add the query listed in \ref{lst:augmented_prompt} for each augmented document. We use the top response generated by the LLMs, with the temperature set to zero to ensure a more deterministic and reliable response~\cite{ouyang2023llm}.

\subsubsection{\textbf{LLM Baselines}}
The following LLM baselines are chosen due to their recency, state-of-the-art, and various architectures: \emph{GPT-3.5-Turbo}\footnote{\url{https://platform.openai.com/docs/models/gpt-3-5-turbo}}, \emph{GPT-4o}\footnote{\url{https://openai.com/index/hello-gpt-4o/}}, \emph{Mistral 8x22B Instruct}\footnote{\url{https://mistral.ai/news/mixtral-8x22b/}}, and \emph{Llama 3.1 405B Instruct} \cite{dubey2024llama}.
These models represent a range of capabilities and architecture designs, from the highly versatile and widely adopted \emph{GPT} series to the specialized mixture of experts (MoE) approach in Mistral and the efficient Llama model.
We also diversified the open and closed-source models, where the \emph{GPTs} are the two closed models, and Mistral and Llama are the open-sourced models.

\input{table/zeroshot}

\subsection{\textbf{Evaluation Metrics}}
\label{sec:eval}
We evaluate the generated test cases both statically and dynamically, utilizing the following commonly used metrics. \respto{7-3} \add{We follow and adopt the metrics used in previous studies to quantitatively assess the performance of the generated unit tests \cite{shin2024domain, alagarsamy2024a3test, schafer2023empirical}.}
\begin{enumerate}
    \item \textbf{Parse Rate} (abbr. PR\%): The AST parsability to check the syntactical correctness of the generated tests. The parse rate is calculated at the test suite level since we prompt the models to generate directly runnable, standalone tests.
    \item \textbf{Execution Rate} (abbr. EX\%): For parsable test suites, we execute them to assess correctness, computing the pass rate as the proportion of tests that run without errors.
    \item \textbf{Pass Rate} (abbr. PS\%): Since not all the executable tests pass, we also report the pass rate of the test cases, calculated as the number of passing tests divided by the total number of executable tests.

    \item \textbf{Line Coverage} (abbr. CV): We calculate line coverage to evaluate the test adequacy of the generated tests. 
    Specifically, after executing a test suite for the API, we calculate the target classes' coverage.
    \item \textbf{Bug Detectability} (abbr. BD): This metric evaluates a unit test's ability to identify faults. We count the number of true bugs revealed by the generated unit test cases.
\end{enumerate}
Note that these metrics are intentionally designed with different granularities to assess the test cases comprehensively. The parse rate is evaluated at the level of the test suite code generated per API. Given that multiple unit test cases are defined within the test suite, we calculate the execution and pass rates at the method level. We assess class-level line coverage, as the code within the class and the API method are highly cohesive. Lastly, BD is calculated at the unit test (method) level.

%% file: figure/overall_fig.tex
\begin{figure}[t!]
\centering
\includegraphics[width=\linewidth]{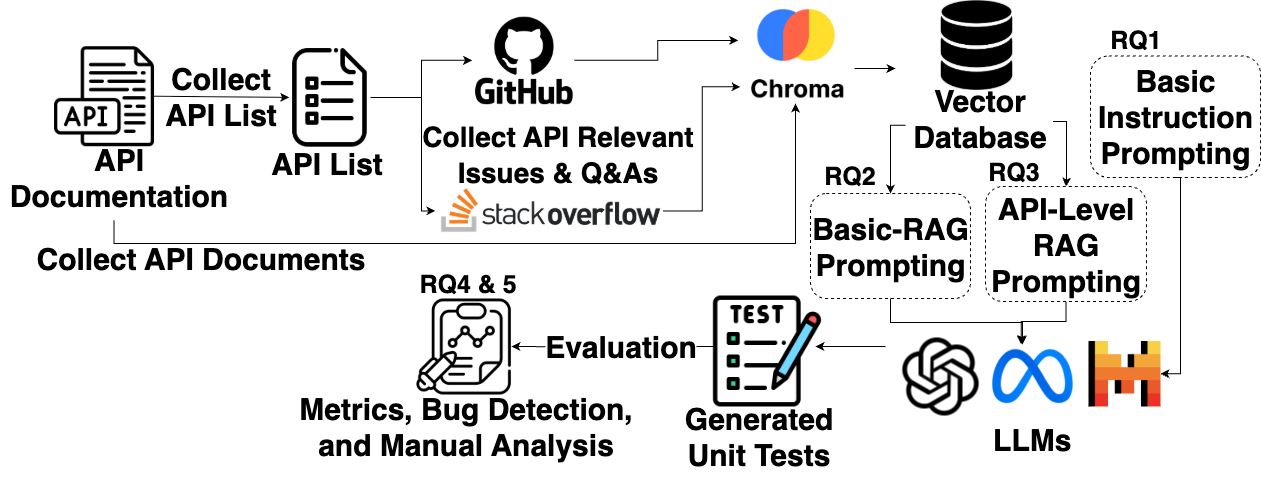}
\vspace{-0.25in}
\caption{\respto{8-1} \respto{8-4} \add{Overview of the proposed method.}}
\label{fig:overall_fig}
\end{figure}

%% file: table/data.tex
\begin{table}[t!]
\centering
\caption{Statistics of collected data.} 
\vspace{-0.1in}
\label{tab:data}
\resizebox{0.7\linewidth}{!}{
\begin{tabular}{|c|c|c|c|}
\hline
\textbf{Project} & \textbf{\# of API Docs} & \textbf{\# of Issues} & \textbf{\#of Q\&As} \\ \hline
\textbf{tf} & 638 & 5,317 & 8,254 \\ \hline
\textbf{torch} & 327 & 7,146 & 5,036 \\ \hline
\textbf{sklearn} & 409 & 4,274 & 7,946 \\ \hline
\textbf{jax} & 257 & 1,708 & 326 \\ \hline
\textbf{xgboost} & 237 & 796 & 464 \\ \hline
\end{tabular}}
\end{table}

%% file: figure/zeroshot_prompt.tex
\lstset{ 
  basicstyle=\ttfamily\tiny,        
  breaklines=true,                 
  commentstyle=\color{brown},    
  firstnumber=1,                
  frame=single,	                   
  keepspaces=true,                 
  keywordstyle=\color{blue},       
  numbers=left,                    
  numbersep=7pt,                   
  numberstyle=\tiny\color{orange}, 
  rulecolor=\color{black},         
  showspaces=false,                
  showstringspaces=false,          
  showtabs=false,                  
  stepnumber=1,                    
  stringstyle=\color{blue},     
  tabsize=2,	                   
  title=\lstname                   
}
\begin{figure}[t!]
\centering
\begin{lstlisting} 
basic_instruction_prompt = [
    {"role": "system",
     "content": f"You are a unit test suite generator for {ML_LIB} library."},
    {"role": "user",
     "content": f"Generate a python unit test case to test the functionality of {API_NAME} API in {ML_LIB} library with maximum coverage. Only create new tests if they cover new lines. Generate test suite using unittest library so it can be directly runnable (with the necessary imports and a main function)."
]
\end{lstlisting}
\vspace{-0.2in}
\caption{The template used for basic instruction prompting.}
\label{lst:zeroshot_prompt}
\end{figure}

%% file: figure/augmented_prompt.tex
\lstset{basicstyle=\ttfamily\tiny,breaklines=true}
\begin{figure}[t!]
\centering
\begin{lstlisting} 
final_prompt = f"{instruction_prompt} Use the following documents (surrounded by
 @@@) to make the test case more compilable and passable and cover more lines.
 @@@ Doc_1: {doc1) @@@\n @@@ Doc_2: {doc2} @@@\n @@@ Doc_3: {doc3} @@@"
\end{lstlisting}
\vspace{-0.2in}
\caption{Additional prompt used for augmented generation.}
\label{lst:augmented_prompt}
\end{figure}

%% file: table/zeroshot.tex
\begin{table*}[t!]
\centering
\caption{\respto{8-3} \respto{8-5} \add{RQ1: Result of basic instruction prompting. The first row denotes the five libraries. The second row denotes the metrics. PR\% denotes parse rate, EX\% denotes execution rate, PS\% denotes pass rate, and CV denotes code coverage. Bold denotes the highest CV among the models.}}
\vspace{-0.1in}
\label{tab:zeroshot}
\setlength{\tabcolsep}{3pt}
\resizebox{\textwidth}{!}{
\begin{tabular}{c|cccc||cccc||cccc||cccc||cccc|}
\cline{2-21}
 & \multicolumn{4}{c||}{\textbf{tf}} & \multicolumn{4}{c||}{\textbf{torch}} & \multicolumn{4}{c||}{\textbf{sklearn}} & \multicolumn{4}{c||}{\textbf{xgb}} & \multicolumn{4}{c|}{\textbf{jax}} \\ \hline
\multicolumn{1}{|c|}{\textbf{FM}} & \multicolumn{1}{c|}{\textbf{PR\%}} & \multicolumn{1}{c|}{\textbf{EX\%}} & \multicolumn{1}{c|}{\textbf{PS\%}} & \textbf{CV} & \multicolumn{1}{c|}{\textbf{PR\%}} & \multicolumn{1}{c|}{\textbf{EX\%}} & \multicolumn{1}{c|}{\textbf{PS\%}} & \textbf{CV} & \multicolumn{1}{c|}{\textbf{PR\%}} & \multicolumn{1}{c|}{\textbf{EX\%}} & \multicolumn{1}{c|}{\textbf{PS\%}} & \textbf{CV} & \multicolumn{1}{c|}{\textbf{PR\%}} & \multicolumn{1}{c|}{\textbf{EX\%}} & \multicolumn{1}{c|}{\textbf{PS\%}} & \textbf{CV} & \multicolumn{1}{c|}{\textbf{PR\%}} & \multicolumn{1}{c|}{\textbf{EX\%}} & \multicolumn{1}{c|}{\textbf{PS\%}} & \textbf{CV} \\ \hline
\multicolumn{1}{|c|}{\textbf{4o}} & \multicolumn{1}{c|}{100.00} & \multicolumn{1}{c|}{89.17} & \multicolumn{1}{c|}{81.29} & \textbf{39.08} & \multicolumn{1}{c|}{100.00} & \multicolumn{1}{c|}{91.09} & \multicolumn{1}{c|}{84.49} & \textbf{25.66 }& \multicolumn{1}{c|}{100.00} & \multicolumn{1}{c|}{92.51} & \multicolumn{1}{c|}{80.22} & \textbf{68.19} & \multicolumn{1}{c|}{100.00} & \multicolumn{1}{c|}{45.70} & \multicolumn{1}{c|}{33.33} & \textbf{56.93} & \multicolumn{1}{c|}{100.00} & \multicolumn{1}{c|}{89.38} & \multicolumn{1}{c|}{83.95} & \textbf{34.45} \\ \hline
\multicolumn{1}{|c|}{\textbf{3.5T}} & \multicolumn{1}{c|}{99.04} & \multicolumn{1}{c|}{71.84} & \multicolumn{1}{c|}{60.58} & 36.72 & \multicolumn{1}{c|}{93.20} & \multicolumn{1}{c|}{88.01} & \multicolumn{1}{c|}{76.40} & 25.43 & \multicolumn{1}{c|}{96.45} & \multicolumn{1}{c|}{91.41} & \multicolumn{1}{c|}{74.48} & 59.76 & \multicolumn{1}{c|}{99.28} & \multicolumn{1}{c|}{64.08} & \multicolumn{1}{c|}{49.51} & 48.12 & \multicolumn{1}{c|}{100.00} & \multicolumn{1}{c|}{81.75} & \multicolumn{1}{c|}{77.78} & 33.14 \\ \hline
\multicolumn{1}{|c|}{\textbf{MS}} & \multicolumn{1}{c|}{95.19} & \multicolumn{1}{c|}{42.73} & \multicolumn{1}{c|}{32.04} & 37.05 & \multicolumn{1}{c|}{95.15} & \multicolumn{1}{c|}{64.19} & \multicolumn{1}{c|}{50.68} & 18.29 & \multicolumn{1}{c|}{92.39} & \multicolumn{1}{c|}{69.67} & \multicolumn{1}{c|}{54.38} & 58.49 & \multicolumn{1}{c|}{95.65} & \multicolumn{1}{c|}{38.46} & \multicolumn{1}{c|}{25.34} & 51.52 & \multicolumn{1}{c|}{97.92} & \multicolumn{1}{c|}{75.98} & \multicolumn{1}{c|}{61.57} & 33.01 \\ \hline
\multicolumn{1}{|c|}{\textbf{LM}} & \multicolumn{1}{c|}{80.77} & \multicolumn{1}{c|}{50.28} & \multicolumn{1}{c|}{43.50} & 32.84 & \multicolumn{1}{c|}{85.44} & \multicolumn{1}{c|}{79.09} & \multicolumn{1}{c|}{71.52} & 17.43 & \multicolumn{1}{c|}{64.97} & \multicolumn{1}{c|}{80.00} & \multicolumn{1}{c|}{70.00} & 22.78 & \multicolumn{1}{c|}{94.93} & \multicolumn{1}{c|}{55.09} & \multicolumn{1}{c|}{55.56} & 16.72 & \multicolumn{1}{c|}{87.50} & \multicolumn{1}{c|}{66.67} & \multicolumn{1}{c|}{55.56} & 30.95 \\ \hline
\end{tabular}
}
\end{table*}

%% file: sections/4.results.tex
\section{Experiment Design and Results}
\label{sec:results}

\subsection{RQ1: Performance of LLMs with Basic Instruction Prompting}
\noindent \textbf{Design.}
In this RQ, 
we prompt the models to generate unit test cases for each API without augmenting them with external sources. We extract only the Python code part from the response, as LLMs tend to generate verbose natural language explanations. 

We parse the generated test suite using the \emph{ast} built-in module and calculate the parse rate. Then, we execute the ones that are parsable. We then parse the log to identify unit test cases that result in errors or failures, thereby calculating the execution and pass rates. For line coverage, we use coverage.py\footnote{\url{https://coverage.readthedocs.io/en/latest/index.html}} to obtain the coverage report. We calculate the coverage score of the class that defines the API to assess the coverage of the affected class. 

It is important to note that these coverage reports do not explore the library's back-end implementation, such as the C++ implementation, which many of these libraries rely on for faster computation. Investigations into the back-end implementation must incorporate each library's specialized testing framework, which requires different expertise, which we leave for future work.

\noindent \textbf{Result.}
Table \ref{tab:zeroshot} presents the results of LLM-generated unit tests across five projects. Most SOTA LLMs can generate syntactically correct unit test cases, with parse rates ranging from 65\% to 100\%. Additionally, most LLMs could generate a considerable number of executable unit test cases, with some exceptions.

This was mainly due to the complex unit tests it generated, which required references to unimported packages, causing them to fail. Similarly, \emph{Mistral} failed to generate executable test cases in certain instances, with a minimum execution rate of 38\%. This failure is attributed to the model's inability to generate directly runnable test suites from the prompt. \emph{Mistral} tends to produce a sequence of unstructured code fragments and natural language, including unnecessary elements such as the code base of the actual API and its explanation. These individual code pieces lack the necessary structure and dependencies, such as the main class, unit test functions, and imports, leading to low executability.

\emph{GPT-4o} consistently achieved the best overall results, ranking first in parse rate and coverage. It also achieved the highest execution rate, except for xgb (3rd). It also ranked best in pass rate, except for xgb (2nd). \emph{GPT-3.5-turbo} and \emph{Mistral} generally ranked in the middle; \emph{GPT-3.5-turbo} being second three times and \emph{Mistral} twice for code coverage. Both models demonstrated good parse rates and relatively high execution and pass rates.


\respto{5-4} \add{\emph{Llama}, on the other hand, had the lowest coverage scores, placing fourth across all projects. This was due to its outputs' low parsability, often mixing code snippets with natural-language descriptions (that prevent automated post-processing from working correctly), and low executability (since it failed to produce standalone, runnable test code). However, the unit test cases generated by Llama appeared valid and appropriate for the target APIs. With a prompt carefully constraining its freedom or a sophisticated parser, Llama shows strong potential to generate effective unit tests.}

We also observed that syntactical and execution correctness do not always correlate with higher coverage. Although more tests can aid in testing different program states, additional cases may result in testing similar or identical parts of the code. For instance, \emph{Mistral} had lower scores in parse, execution, and pass rates for xgb compared to \emph{GPT-3.5-turbo}. Despite executing fewer test cases (102 vs. 106), \emph{Mistral} achieved higher code coverage (52\% vs. 48\%).

While it is important to measure the syntactical and execution correctness of generated unit tests, from a software testing perspective, the code coverage of the test suites is ultimately more important. Therefore, in subsequent RQs, we will further investigate the effectiveness of RAG on code coverage. 

\mybox{\textbf{Answer to RQ1:} The state-of-the-art LLMs have demonstrated promising results in generating syntactic and executable unit test cases. \emph{GPT-4o} showed the most promising results in most aspects. From the initial study, we also observe that syntactical and execution correctness do not always lead to higher code coverage, suggesting further investigation into code coverage.}

\input{table/basic_a}

\subsection{RQ2: Impact of External Sources in Basic RAG}
\noindent \textbf{Design.}
This RQ aims to investigate how different external sources, namely API documentation, GitHub issues, and StackOverflow Q\&As, can enhance the Basic RAG in generating effective and practical unit tests. To determine which documents are most effective in complementing test case generation, we construct four vector databases: (1) API documentation only, (2) GitHub issues only, (3) StackOverflow Q\&As only, and (4) a combination of all three. 

Since Basic RAG does not differentiate between document types, all documents from the five projects are aggregated into each of the four databases. Similar to the previous RQ, we evaluate the parse rate, execution rate, pass rate, and line coverage.

To further investigate the impact of Basic RAG on code coverage performance, we undertake the following steps: (1) count the wins of each RAG approach versus basic instruction prompting, (2) count the number of wins for each Basic RAG approach against each other, and (3) perform a pairwise ranking of the Basic RAG approaches with the four databases and the basic instruction prompting using the Friedman tests \cite{demvsar2006statistical} among basic instruction prompting, Basic RAG combined, Basic RAG API documents, Basic RAG GitHub issues, and Basic RAG StackOverflow Q\&As. 

\noindent \textbf{Result.}
Table \ref{tab:basic_a} presents the results of LLMs in generating unit tests using the Basic RAG. For comparison, we also report the results of basic instruction prompting. Overall, we observe that employing Basic RAG can enhance code coverage compared to basic instruction prompting. However, the augmentation of documents appears to negatively impact the parse rate for certain models, such as \emph{GPT-3.5-turbo} and \emph{Mistral}. This is attributed to the fact that these relatively smaller models are disrupted by the lengthy augmented documents, causing them to deviate from the initial prompt to ``generate directly runnable test suites'' and produce unparsable code \respto{5-5} \add{\cite{hosseini2025efficient}.}

We report the win count on code coverage for each approach, as depicted in Figure \ref{fig:win_counts}a and \ref{fig:win_counts}c. In a total of 20 cases (5 projects $\times$ 4 LLMs), the combined Basic RAG outperformed basic instruction prompting 13 times, the API document Basic RAG 12 times, the GitHub issues Basic RAG 14 times, and the StackOverflow Q\&As Basic RAG 11 times. Contrary to our assumption that combined RAGs would outperform individual RAGs due to the inclusion of all documents, the results did not consistently support this hypothesis. When comparing the combined Basic RAG to the individual RAGs (Figure \ref{fig:win_counts}c), the combined Basic RAG outperformed the API documents 11 times, the GitHub issues 10 times, and the StackOverflow Q\&As 12 times.
From the Friedman ranking test, we obtained the following average rankings: (1) Basic combined: 2.7, 2) Basic RAG GitHub issues: 2.7, 3) Basic RAG API documents: 3.05, 4) Basic RAG StackOverflow Q\&As: 3.05, 5) Basic instruction prompting: 3.5. The p-value from the Friedman test was 0.4809, indicating no statistical significance of the rankings (p-value > 0.05).


The results indicate that GitHub issues were equally effective as the combined Basic RAG in generating unit test cases, both having the same average ranking of 2.7. A possible reason for this could be that GitHub issues provide detailed code examples for teams to replicate their bugs, and through multiple comments, they offer detailed mitigation strategies. Further analysis through manual assessment will be discussed in RQ5.

\mybox{\textbf{Answer to RQ2:} The application of Basic RAG improves the generation of unit tests, particularly in code coverage. However, the statistical significance of ordering was not significant using Basic RAGs.}

\input{figure/win_counts}

\subsection{RQ3: Impact of External Sources in API-Level RAG}
\noindent \textbf{Design.}
Similar to RQ2, this research question investigates the effects of different external sources within an API-level RAG setting. Utilizing the available mapping information that links documents to each API, we construct a dedicated vector database for each API. As with the previous RQs, we calculate the four metrics, count the wins for each RAG approach, and perform the Friedman tests on the code coverage for the four API RAGs versus basic instruction prompting. Additionally, we conduct a Friedman test comparing the basic and API RAG approaches, resulting in nine rank comparisons. 

\input{table/api_a}

\noindent \textbf{Result.}
Table \ref{tab:api_a} presents the results of LLMs in generating unit test cases using the API-level RAG. For comparative purposes, basic instruction prompting results are also included. Consistent with the findings in RQ2, it is evident that the API-level enhances performance in terms of code coverage compared to basic instruction prompting. A similar decline in parse rate was observed for \emph{GPT-3.5-turbo} and \emph{Mistral}, attributable to the same underlying factors. 

Figures \ref{fig:win_counts}b and \ref{fig:win_counts}d illustrate the win counts of API-level RAG versus basic instruction prompting within each API-level RAG. Out of 20 cases, the combined API-level RAG outperformed basic instruction prompting in 15 instances, the API document API-level RAG in 13 instances, the GitHub issues API-level RAG in 14 instances, and the StackOverflow Q\&As API-level RAG in 12 instances. The win count for API-level RAG versus basic instruction prompting was higher compared to Basic RAG versus basic instruction prompting, with two additional wins for the combined approach, one additional win for the API document approach, an equal number of wins for the GitHub issues approach, and one additional win for the StackOverflow Q\&As approach.

Furthermore, the number of wins for the combined API-level RAG versus the individual RAGs was also analyzed. Similar to the Basic RAG results, the combined API-level RAG outperformed the API documents 17 times, GitHub issues 8 times, and StackOverflow Q\&As 13 times. Compared to Basic RAG, the win counts increased by six for API documents and one for StackOverflow Q\&As, while decreasing by two for GitHub issues. These results suggest a trend similar to Basic RAG, with the combined RAG outperforming the other RAGs except for the GitHub issues RAG.


Additionally, we conducted the Friedman ranking test on the line coverage from API-level RAG and basic instruction prompting and the results yielded the following rankings: 1) API RAG GitHub issues: 2.3, 2) API RAG combined: 2.35, 3) API RAG StackOverflow Q\&As: 3.0, 4) API RAG API documentation: 3.65, 5) Basic instruction prompting: 3.7. This indicates that for API-level RAG, GitHub issues were more effective than the combined version, a trend also observed in the previous research question. The p-value from the Friedman test was 0.0056, indicating the significance of the rankings (p-value < 0.05).
\respto{5-3} \add{To demonstrate the magnitude of the difference between the methods, we have augmented the Friedman test with Kendall's W \cite{kendall1939problem}. The effect size for this ranking was 0.38, indicating a moderate effect size.}

Finally, we conducted the Friedman ranking test for all combined approaches, which yielded the following rankings: 
1) API RAG GitHub issues: 3.4, 2) API RAG combined: 3.7, 3) Basic RAG combined: 4.9, 4)Basic RAG GitHub issues: 4.9, 5) API RAG StackOverflow Q\&As: 4.9, 6) Basic RAG StackOverflow Q\&As: 5.55, 7) Basic RAG API Documentation: 5.75, 8) API RAG API Documentation: 5.8, 9) basic instruction prompting: 6.2.
These rankings indicate that API RAG on GitHub issues was the most effective. Additionally, API RAG generally achieved better rankings than Basic RAG. The p-value from the Friedman test was 0.0127, confirming the significance of the rankings (p-value < 0.05). \respto{5-3} \add{The effect size using Kendall's W for this ranking was 0.23, indicating a small effect size.}
\vspace{-0.1in}
\mybox{\textbf{Answer to RQ3:} API-level RAG has been shown to improve unit test generation, particularly in enhancing code coverage, outperforming Basic RAG. Notably, API-level RAG incorporating GitHub issues achieves the best performance, with statistically significant improvements.}

\subsection{RQ4: Bug Detection}
\noindent \textbf{Design.}
This research question investigates the effectiveness of the generated tests in detecting real-world bugs. Since no established bug benchmark exists for ML/DL APIs, we manually evaluate bug detectability. 
\respto{3-1a} \add{For this experiment, we focus on test cases generated by \emph{GPT-4o}, the best-performing model, on the same sets of APIs as the previous RQs.} Given that GitHub issues and StackOverflow Q\&As often highlight common bugs that developers encounter, there's a potential risk of information leakage. Thus, we compare the API documentation RAG approach at the API level against basic instruction prompting. While the previous RQs prioritized test effectiveness in terms of coverage, here we explicitly direct the models to generate test cases aimed at revealing bugs by including the following instruction: ``Your ultimate goal is to generate test cases that reveal bugs in this API.''
For the evaluation, we only consider generated test cases that exhibit failure for bug detection.
We then analyze the failing test cases and their execution logs to determine whether the test assumptions align with the API documentation. If so, the failure is likely caused by an issue in the API's source implementation.
Two authors manually label the detected failures, resolving any disagreements through discussion \cite{shen2023characterization, li2023they}. For the confirmed true positives, we search the issue repository for similar bugs previously and only report the new ones.
The links to the reported issues are included in the replication package.

\input{table/BD_dist}

\noindent \textbf{Result.}
\respto{3-1a} \add{Table \ref{tab:bd_dist} summarizes the bug‐detection results for our generated unit tests. We first report the total number of APIs (test class-level) and the total number of unit test cases generated (test method-level). We then present the number of failing tests, which are the ones of our interest. Next are the execution and pass rates, using the same definitions as the previous RQs. And finally, the bug detectability, where the failure reveals a true positive of a real bug.} \respto{5-6b} \add{From the results, we observe that most unit tests still pass (average 69.5\%) despite our prompts to generate bug-revealing tests. Also, out of the failed tests, only a few reveal real bugs. This suggests that the libraries robustly handle common exceptions and that it is very challenging for the LLMs to generate unit tests that reveal true bugs.}

Using the API Documentation RAG approach, we identified a total of \respto{3-1a} \add{19 true bugs from the five libraries. In contrast, basic instruction prompting uncovered 9 true bugs. In total, the tests detected 28 bugs. Out of them, three were redundant tests (either revealing the same bug or similar tests generated from both basic prompts and API doc RAG), and one of them had already been reported by someone else. So in total, 24 new bugs were reported to their respective GitHub issue repository.} \respto{6-1} \add{At the time of submission, the developers confirmed ten as bugs, four as not a bug, and the rest (ten) are awaiting confirmation. The progress of this can be followed by the links that we provide in the replication package.}

\input{figure/bug_ex}

Figure \ref{lst:bug_ex} illustrates a unit test generated by \emph{GPT-4o} using the API Documentation RAG approach, which reveals a bug. The API documentation specifies that only boolean types should be used as mask values; however, the current implementation doesn't enforce or validate this, leading to an assertion error.

Most detected errors stemmed from inconsistencies between the API implementation and its documentation or from inadequate validation mechanisms that failed to raise the expected errors, which aligns with previous findings \cite{harzevili2023automatic, harzevili2024checker}.
Since the prompt explicitly instructed the models to generate test cases for bug detection, many of the tests pushed the API into invalid states and asserted specific error conditions, functioning similarly to fuzz testing.
API documentation provides detailed specifications regarding expected input types and values, which helped the API Documentation RAG approach generate more comprehensive test cases to check invalid inputs compared to basic instruction prompting. As a result, the API Documentation RAG was more effective in uncovering bugs, as observed in the example. We can observe that not all failed tests can detect bugs. 
\respto{5-10} \add{This issue arises from the model's hallucinations, which lead to an incorrect understanding of the API's intended behaviour, generating incorrect assertions, i.e., test cases that do not align with the actual API specification \cite{taherkhani2024valtest}.}

\mybox{\textbf{Answer to RQ4:} LLM-based unit test case generation can detect real-world bugs. Using API Documentation RAG prompting, we found it detected more bugs (19 vs. 9) by generating test cases more aligned with the API documentation. These bugs often stemmed from implementation inconsistencies or insufficient validation mechanisms. We reported 24 new bugs in their respective issue repositories.}

\input{figure/venn}
\input{figure/bar}

\subsection{RQ5: Qualitative Analysis of RAG}
\noindent \textbf{Design.}
This RQ aims to investigate how different knowledge sources support the generation of more comprehensive and effective tests. 
\respto{3-1b} \add{To maintain statistical significance while keeping manual analysis feasible, we randomly sampled 250 APIs out of the total 694 (the minimum threshold for 95\% confidence with a 5\% margin of error is 248). We sampled equal distribution from each library, selecting 75 from tf, 37 from torch, 70 from sklearn, 18 from jax, and 50 from xgb. All evaluations focus on tests generated by \emph{GPT-4o}, with the four API-level RAGs and the basic instruction as our baseline approaches.}

Before the manual evaluation, we first identify 
unique lines covered by each method but not by others. We then analyze the types of information contained in the retrieved documents and how they influenced or guided the model in generating tests that exercised these unique lines. 

\respto{4-1} \add{Next, we categorize the retrieved documents according to the types of information they provide to the LLM, such as usage examples, parameter constraints, or error-handling patterns, and analyze their specific roles and impact in guiding the model's test generation. First, we started with initial categories based on our knowledge of the three types of documents: bug-inducing (issues), standard use cases (documentation), and unique/edge use cases (Q\&As). Then we performed open‑coding thematic analysis, clustered codes through a systematic taxonomy development method, and built a concise code-book \cite{guest2011applied}. Two participants independently applied it across all documents, and two more participants were involved to resolve disagreements through consensus. The following is the final list of seven information categories contributed by the retrieved documents: 
\begin{enumerate}
    \item Bug-inducing \cite{catolino2019not, hong2013effective}: Documents under this category often reveal actual defects or incorrect software behaviour, helping the model learn faulty usage patterns or edge cases. 
    \item Standard use cases \cite{felderer2014taxonomy, wang2020automatic}: Documents under this category often provide clear and concise examples that demonstrate typical API interactions, helping guide the model toward generating valid and representative test cases. 
    \item API usage sequences \cite{zhang2022efficient, niu2017api}:  Documents under this category often illustrate multi-step workflows and sequences of dependent API calls, helping the model understand typical API invocation patterns in realistic contexts. 
    \item Unique/complex input/parameters \cite{guo2024optimal, sen2005cute}: Documents under this category often showcase advanced or edge-case input configurations and parameter settings, enabling the model to generate tests that explore uncommon or corner-case behaviours. 
    \item Exception handling patterns \cite{zhang2012amplifying, de2017studying}: Documents under this category often explain proper error handling strategies and the management of expected failure conditions, guiding the model to generate more robust and fault-tolerant tests. 
    \item Integration patterns \cite{schwarz2025taxonomy, saadatmand2019integrationdistiller}: These documents describe how APIs interoperate within larger systems or interact with external libraries, helping the model construct tests that reflect real-world integration scenarios. 
    \item Performance considerations \cite{costa2020taxonomy, trubiani2011detection}: Documents in this category focus on efficiency, scalability, and best practices for resource usage, enabling the generation of tests that capture performance-critical behaviours. 
\end{enumerate}
Note that the number of categories doesn't always match the number of documents per instance, as a document could have multiple (or even no) parts that influence the test generation.
Also, that basic instruction was only used for set deduction to calculate unique coverage, as they do not use any external documents for test generation.
All the artifacts that were produced in this process can be found in our replication package.}

\input{table/manual_category}

\noindent \textbf{Result.}
\respto{3-1b} \respto{4-1}
\add{We first show how each RAG source affects the unique line coverage. Figure~\ref{fig:venn} presents the number of APIs for which each approach achieves unique line coverage, reflecting breadth-wise API coverage. Figure~\ref{fig:bar} shows the depth-wise line coverage, measured as the average number of uniquely covered lines per approach. 
From these figures, we can observe that GitHub issues have the most APIs with unique lines and the average number of lines per API, aligning with the previous RQs. We can also observe that most APIs are covered (97) when using StackOverflow Q\&As, GitHub Issues, and the combined approach. However, if we look at average unique lines covered, API documentation was in second place, not StackOverflow. With the numerous demonstrations of API sequences, StackOverflow gave wide coverage in terms of API numbers. However, with the depth of the information regarding all the parameters provided by the API documentation, the absolute number of lines covered was higher. Basic instruction showed unique coverages, showing 11 APIs that only basic instruction had unique coverage. It also had relatively high average line coverage (7.9), having a higher count than combined and StackOverflow, indicating that higher coverage can be achieved when we utilize them complementarily.\\
\indent Table \ref{tab:category} shows the overall results from our manual analysis, the number of each category identified per RAG approach. We can observe that GitHub issues had the most identified documents that influenced tests with unique coverage, with the most number of categories for four out of seven. Combined RAG was the next RAG approach with the most identified document categories and with the most standard use cases. Although it didn't cover the next most unique lines, it covered the next most APIs in breadth. API Documentation had the least amount of categorized documents and the least variety, due to its fixed style of documentation. The example code and tutorials contributed to standard use cases and API usage sequences, and the detailed parameter information contributed to unique/edge use cases (unique input/parameters). 
StackOverflow Q\&As exhibited the most varied sequences of API usage and integration patterns, as both questioners and answerers contribute custom code snippets that also use other third-party libraries for integration. In contrast, the API documentation featured a high number of unique and complex input and parameter categories, providing comprehensive details about all parameters. This information guided LLMs in generating unique parameters that are often set as default in standard use cases. In theory, the combined RAG could incorporate advantages from all three sources. However, it shows that applying either source benefits in unique coverage by focusing on its strengths further.}

\mybox{\textbf{Answer to RQ5:} From our manual analysis, we identify seven types of information in documents that aid test generation. We also find GitHub issues cover more lines by providing diverse and unique testing knowledge.}

\add{\subsection{Discussions}}
\respto{7-4b} \respto{7-6}
\add{
\subsubsection{Key Insights}
Our experimental results show that RAG significantly enhances LLM-based test generation, especially for complex and domain-specific tasks. Although pre-training data may contain relevant documentation, LLMs often struggle to recall specific details or stay up-to-date with API changes. This limitation arises from its static and incomplete parametric memory \cite{jain2024mitigating, chen2025llms}. RAG addresses this by providing up-to-date, API-specific snippets at generation time, highlighting edge cases from community-reported issues the model may have forgotten. Rather than supplying entirely new data, RAG grounds generation in precise, relevant context. This reduces ambiguity, limits hallucinations, leading to measurable gains in coverage and bug detection.
Our study provides empirical evidence: RAG achieved a 6.5\% point increase in code coverage and identified 10 additional real-world bugs, which is 111\% more than without RAG.
\subsubsection{Implications for Research and Practice}
For researchers, the strong performance of API-level RAG, especially when sourcing from GitHub issues, highlights the value of fine-grained, domain-specific retrieval strategies. Future work should explore automated document selection, hybrid sparse-dense indexing, and extending evaluation to mutation or property-based coverage.
In practice, teams aiming to improve code coverage and bug detection should consider integrating an API-level RAG pipeline. For best results, build a lightweight vector store of GitHub issue texts and query it alongside official docs. Budget permitting, GPT-4o is the preferred generator; for cost-sensitive settings, \emph{GPT-3.5-Turbo} may suffice, but with modest performance trade-offs. 
\subsubsection{Cost, Challenges, and Limitations}
Building and maintaining per-API vector databases incurs storage overhead and query latency, and must be refreshed as APIs evolve. Smaller LLMs (e.g., \emph{Mistral, Llama}) faced parse rate declines when handling lengthy retrieved documents, indicating a trade-off between context richness and model capacity. Our requirement of at least 10 documents per API excludes less-popular functions, limiting coverage of newly introduced APIs.}

%% file: table/basic_a.tex
\begin{table*}[t!]
\caption{RQ2: Result of LLMs for using basic RAG prompting. BL denotes the baselines: basic instruction (BI), combined (CB), GitHub issues (GH), and StackOverflow Q\&As (SO). In the second row, PR\% denotes parse rate, EX\% denotes execution rate, PS\% denotes pass rate, and CV denotes coverage. Bold denotes the highest CV among the prompting strategies.}
\vspace{-0.1in}
\label{tab:basic_a}
\setlength{\tabcolsep}{3pt}
\resizebox{\textwidth}{!}{
\begin{tabular}{cc|cccc||cccc||cccc||cccc||cccc|}
\cline{3-22}
\multicolumn{1}{l}{} & \multicolumn{1}{l|}{} & \multicolumn{4}{c||}{\textbf{tf}} & \multicolumn{4}{c||}{\textbf{torch}} & \multicolumn{4}{c||}{\textbf{sklearn}} & \multicolumn{4}{c||}{\textbf{xgb}} & \multicolumn{4}{c|}{\textbf{jax}} \\ \hline
\multicolumn{1}{|c|}{\textbf{LLM}} & \textbf{BL} & \multicolumn{1}{c|}{\textbf{PR\%}} & \multicolumn{1}{c|}{\textbf{EX\%}} & \multicolumn{1}{c|}{\textbf{PS\%}} & \textbf{CV} & \multicolumn{1}{c|}{\textbf{PR\%}} & \multicolumn{1}{c|}{\textbf{EX\%}} & \multicolumn{1}{c|}{\textbf{PS\%}} & \textbf{CV} & \multicolumn{1}{c|}{\textbf{PR\%}} & \multicolumn{1}{c|}{\textbf{EX\%}} & \multicolumn{1}{c|}{\textbf{PS\%}} & \textbf{CV} & \multicolumn{1}{c|}{\textbf{PR\%}} & \multicolumn{1}{c|}{\textbf{EX\%}} & \multicolumn{1}{c|}{\textbf{PS\%}} & \textbf{CV} & \multicolumn{1}{c|}{\textbf{PR\%}} & \multicolumn{1}{c|}{\textbf{EX\%}} & \multicolumn{1}{c|}{\textbf{PS\%}} & \textbf{CV} \\ \hline
\multicolumn{1}{|c|}{\multirow{5}{*}{\textbf{4o}}} & \textbf{BI} & \multicolumn{1}{c|}{100.00} & \multicolumn{1}{c|}{89.17} & \multicolumn{1}{c|}{81.29} & 39.08 & \multicolumn{1}{c|}{100.00} & \multicolumn{1}{c|}{91.09} & \multicolumn{1}{c|}{84.49} & 25.66 & \multicolumn{1}{c|}{100.00} & \multicolumn{1}{c|}{92.51} & \multicolumn{1}{c|}{80.22} & 68.19 & \multicolumn{1}{c|}{100.00} & \multicolumn{1}{c|}{45.70} & \multicolumn{1}{c|}{33.33} & 56.93 & \multicolumn{1}{c|}{100.00} & \multicolumn{1}{c|}{89.38} & \multicolumn{1}{c|}{83.95} & 34.45 \\ \cline{2-22} 
\multicolumn{1}{|c|}{} & \textbf{CB} & \multicolumn{1}{c|}{100.00} & \multicolumn{1}{c|}{85.49} & \multicolumn{1}{c|}{75.29} & 39.12 & \multicolumn{1}{c|}{100.00} & \multicolumn{1}{c|}{89.28} & \multicolumn{1}{c|}{81.00} & 31.21 & \multicolumn{1}{c|}{100.00} & \multicolumn{1}{c|}{88.68} & \multicolumn{1}{c|}{74.55} & 66.52 & \multicolumn{1}{c|}{100.00} & \multicolumn{1}{c|}{60.27} & \multicolumn{1}{c|}{45.21} & 56.06 & \multicolumn{1}{c|}{100.00} & \multicolumn{1}{c|}{85.38} & \multicolumn{1}{c|}{76.32} & 34.59 \\ \cline{2-22} 
\multicolumn{1}{|c|}{} & \textbf{AD} & \multicolumn{1}{c|}{100.00} & \multicolumn{1}{c|}{85.30} & \multicolumn{1}{c|}{76.33} & 39.25 & \multicolumn{1}{c|}{100.00} & \multicolumn{1}{c|}{88.45} & \multicolumn{1}{c|}{83.17} & 18.77 & \multicolumn{1}{c|}{100.00} & \multicolumn{1}{c|}{85.65} & \multicolumn{1}{c|}{74.01} & \textbf{69.52} & \multicolumn{1}{c|}{100.00} & \multicolumn{1}{c|}{59.81} & \multicolumn{1}{c|}{47.91} & \textbf{57.17} & \multicolumn{1}{c|}{100.00} & \multicolumn{1}{c|}{86.89} & \multicolumn{1}{c|}{80.46} & \textbf{35.03} \\ \cline{2-22} 
\multicolumn{1}{|c|}{} & \textbf{GH} & \multicolumn{1}{c|}{100.00} & \multicolumn{1}{c|}{87.58} & \multicolumn{1}{c|}{78.40} & 39.28 & \multicolumn{1}{c|}{100.00} & \multicolumn{1}{c|}{90.50} & \multicolumn{1}{c|}{83.25} & \textbf{31.31} & \multicolumn{1}{c|}{98.98} & \multicolumn{1}{c|}{89.37} & \multicolumn{1}{c|}{75.30} & 67.15 & \multicolumn{1}{c|}{100.00} & \multicolumn{1}{c|}{72.48} & \multicolumn{1}{c|}{54.13} & 55.42 & \multicolumn{1}{c|}{100.00} & \multicolumn{1}{c|}{85.38} & \multicolumn{1}{c|}{79.53} & 34.66 \\ \cline{2-22} 
\multicolumn{1}{|c|}{} & \textbf{SO} & \multicolumn{1}{c|}{100.00} & \multicolumn{1}{c|}{75.83} & \multicolumn{1}{c|}{69.61} & \textbf{39.35} & \multicolumn{1}{c|}{100.00} & \multicolumn{1}{c|}{92.15} & \multicolumn{1}{c|}{86.88} & 25.83 & \multicolumn{1}{c|}{100.00} & \multicolumn{1}{c|}{92.83} & \multicolumn{1}{c|}{80.98} & 64.94 & \multicolumn{1}{c|}{100.00} & \multicolumn{1}{c|}{66.19} & \multicolumn{1}{c|}{52.16} & 55.93 & \multicolumn{1}{c|}{97.92} & \multicolumn{1}{c|}{94.12} & \multicolumn{1}{c|}{88.24} & 34.79 \\ \hline \hline
\multicolumn{1}{|c|}{\multirow{5}{*}{\textbf{3.5T}}} & \textbf{BI} & \multicolumn{1}{c|}{99.04} & \multicolumn{1}{c|}{71.84} & \multicolumn{1}{c|}{60.58} & 36.72 & \multicolumn{1}{c|}{93.20} & \multicolumn{1}{c|}{88.01} & \multicolumn{1}{c|}{76.40} & 25.43 & \multicolumn{1}{c|}{96.45} & \multicolumn{1}{c|}{91.41} & \multicolumn{1}{c|}{74.48} & 59.76 & \multicolumn{1}{c|}{99.28} & \multicolumn{1}{c|}{64.08} & \multicolumn{1}{c|}{49.51} & 48.12 & \multicolumn{1}{c|}{100.00} & \multicolumn{1}{c|}{81.75} & \multicolumn{1}{c|}{77.78} & 33.14 \\ \cline{2-22} 
\multicolumn{1}{|c|}{} & \textbf{CB} & \multicolumn{1}{c|}{95.67} & \multicolumn{1}{c|}{68.71} & \multicolumn{1}{c|}{57.78} & \textbf{37.95} & \multicolumn{1}{c|}{94.17} & \multicolumn{1}{c|}{82.97} & \multicolumn{1}{c|}{71.38} & \textbf{30.95} & \multicolumn{1}{c|}{97.46} & \multicolumn{1}{c|}{78.97} & \multicolumn{1}{c|}{60.82} & 56.35 & \multicolumn{1}{c|}{100.00} & \multicolumn{1}{c|}{52.53} & \multicolumn{1}{c|}{43.04} & 45.61 & \multicolumn{1}{c|}{95.83} & \multicolumn{1}{c|}{76.97} & \multicolumn{1}{c|}{70.39} & 32.80 \\ \cline{2-22} 
\multicolumn{1}{|c|}{} & \textbf{AD} & \multicolumn{1}{c|}{96.63} & \multicolumn{1}{c|}{75.52} & \multicolumn{1}{c|}{65.17} & 36.89 & \multicolumn{1}{c|}{93.20} & \multicolumn{1}{c|}{79.39} & \multicolumn{1}{c|}{68.79} & 18.09 & \multicolumn{1}{c|}{94.42} & \multicolumn{1}{c|}{82.44} & \multicolumn{1}{c|}{66.81} & 55.89 & \multicolumn{1}{c|}{97.10} & \multicolumn{1}{c|}{56.71} & \multicolumn{1}{c|}{40.24} & \textbf{49.65} & \multicolumn{1}{c|}{95.83} & \multicolumn{1}{c|}{69.44} & \multicolumn{1}{c|}{63.89} & \textbf{34.02} \\ \cline{2-22} 
\multicolumn{1}{|c|}{} & \textbf{GH} & \multicolumn{1}{c|}{95.67} & \multicolumn{1}{c|}{69.76} & \multicolumn{1}{c|}{56.80} & 37.22 & \multicolumn{1}{c|}{97.09} & \multicolumn{1}{c|}{84.46} & \multicolumn{1}{c|}{71.28} & 30.92 & \multicolumn{1}{c|}{92.68} & \multicolumn{1}{c|}{74.19} & \multicolumn{1}{c|}{62.37} & 44.51 & \multicolumn{1}{c|}{100.00} & \multicolumn{1}{c|}{55.56} & \multicolumn{1}{c|}{48.77} & 44.80 & \multicolumn{1}{c|}{100.00} & \multicolumn{1}{c|}{75.46} & \multicolumn{1}{c|}{63.19} & 33.66 \\ \cline{2-22} 
\multicolumn{1}{|c|}{} & \textbf{SO} & \multicolumn{1}{c|}{91.35} & \multicolumn{1}{c|}{35.63} & \multicolumn{1}{c|}{26.57} & 35.96 & \multicolumn{1}{c|}{95.15} & \multicolumn{1}{c|}{79.93} & \multicolumn{1}{c|}{67.03} & 25.62 & \multicolumn{1}{c|}{95.43} & \multicolumn{1}{c|}{74.15} & \multicolumn{1}{c|}{60.84} & 53.68 & \multicolumn{1}{c|}{99.28} & \multicolumn{1}{c|}{64.60} & \multicolumn{1}{c|}{50.31} & 41.56 & \multicolumn{1}{c|}{70.83} & \multicolumn{1}{c|}{62.79} & \multicolumn{1}{c|}{50.00} & 33.50 \\ \hline \hline
\multicolumn{1}{|c|}{\multirow{5}{*}{\textbf{MS}}} & \textbf{BI} & \multicolumn{1}{c|}{95.19} & \multicolumn{1}{c|}{42.73} & \multicolumn{1}{c|}{32.04} & 37.05 & \multicolumn{1}{c|}{95.15} & \multicolumn{1}{c|}{64.19} & \multicolumn{1}{c|}{50.68} & 18.29 & \multicolumn{1}{c|}{92.39} & \multicolumn{1}{c|}{69.67} & \multicolumn{1}{c|}{54.38} & \textbf{58.49} & \multicolumn{1}{c|}{95.65} & \multicolumn{1}{c|}{38.46} & \multicolumn{1}{c|}{25.34} & \textbf{51.52} & \multicolumn{1}{c|}{97.92} & \multicolumn{1}{c|}{75.98} & \multicolumn{1}{c|}{61.57} & 33.01 \\ \cline{2-22} 
\multicolumn{1}{|c|}{} & \textbf{CB} & \multicolumn{1}{c|}{74.04} & \multicolumn{1}{c|}{44.66} & \multicolumn{1}{c|}{35.28} & \textbf{37.68} & \multicolumn{1}{c|}{64.08} & \multicolumn{1}{c|}{51.80} & \multicolumn{1}{c|}{35.97} & \textbf{31.21} & \multicolumn{1}{c|}{72.59} & \multicolumn{1}{c|}{63.77} & \multicolumn{1}{c|}{46.15} & 49.32 & \multicolumn{1}{c|}{82.61} & \multicolumn{1}{c|}{45.86} & \multicolumn{1}{c|}{24.81} & 41.35 & \multicolumn{1}{c|}{85.42} & \multicolumn{1}{c|}{72.16} & \multicolumn{1}{c|}{55.11} & 33.53 \\ \cline{2-22} 
\multicolumn{1}{|c|}{} & \textbf{AD} & \multicolumn{1}{c|}{55.77} & \multicolumn{1}{c|}{41.98} & \multicolumn{1}{c|}{27.30} & 37.03 & \multicolumn{1}{c|}{37.86} & \multicolumn{1}{c|}{59.17} & \multicolumn{1}{c|}{54.17} & 17.49 & \multicolumn{1}{c|}{52.79} & \multicolumn{1}{c|}{65.57} & \multicolumn{1}{c|}{53.01} & 42.72 & \multicolumn{1}{c|}{51.45} & \multicolumn{1}{c|}{39.81} & \multicolumn{1}{c|}{26.85} & 42.74 & \multicolumn{1}{c|}{43.75} & \multicolumn{1}{c|}{67.74} & \multicolumn{1}{c|}{50.00} & 31.79 \\ \cline{2-22} 
\multicolumn{1}{|c|}{} & \textbf{GH} & \multicolumn{1}{c|}{82.21} & \multicolumn{1}{c|}{41.37} & \multicolumn{1}{c|}{30.58} & 37.51 & \multicolumn{1}{c|}{63.11} & \multicolumn{1}{c|}{51.77} & \multicolumn{1}{c|}{35.46} & 30.82 & \multicolumn{1}{c|}{76.14} & \multicolumn{1}{c|}{58.33} & \multicolumn{1}{c|}{37.35} & 49.94 & \multicolumn{1}{c|}{86.23} & \multicolumn{1}{c|}{48.20} & \multicolumn{1}{c|}{29.50} & 44.61 & \multicolumn{1}{c|}{89.58} & \multicolumn{1}{c|}{76.27} & \multicolumn{1}{c|}{57.63} & \textbf{35.04} \\ \cline{2-22} 
\multicolumn{1}{|c|}{} & \textbf{SO} & \multicolumn{1}{c|}{75.48} & \multicolumn{1}{c|}{31.40} & \multicolumn{1}{c|}{21.04} & 35.64 & \multicolumn{1}{c|}{86.41} & \multicolumn{1}{c|}{61.00} & \multicolumn{1}{c|}{35.00} & 25.47 & \multicolumn{1}{c|}{59.90} & \multicolumn{1}{c|}{60.49} & \multicolumn{1}{c|}{44.96} & 47.64 & \multicolumn{1}{c|}{57.97} & \multicolumn{1}{c|}{39.60} & \multicolumn{1}{c|}{23.49} & 50.10 & \multicolumn{1}{c|}{75.00} & \multicolumn{1}{c|}{52.50} & \multicolumn{1}{c|}{40.00} & 32.88 \\ \hline \hline
\multicolumn{1}{|c|}{\multirow{5}{*}{\textbf{LM}}} & \textbf{BI} & \multicolumn{1}{c|}{80.77} & \multicolumn{1}{c|}{50.28} & \multicolumn{1}{c|}{43.50} & 32.84 & \multicolumn{1}{c|}{85.44} & \multicolumn{1}{c|}{79.09} & \multicolumn{1}{c|}{71.52} & 17.43 & \multicolumn{1}{c|}{64.97} & \multicolumn{1}{c|}{80.00} & \multicolumn{1}{c|}{70.00} & 22.78 & \multicolumn{1}{c|}{94.93} & \multicolumn{1}{c|}{55.09} & \multicolumn{1}{c|}{55.56} & 16.72 & \multicolumn{1}{c|}{87.50} & \multicolumn{1}{c|}{66.67} & \multicolumn{1}{c|}{55.56} & 30.95 \\ \cline{2-22} 
\multicolumn{1}{|c|}{} & \textbf{CB} & \multicolumn{1}{c|}{86.54} & \multicolumn{1}{c|}{69.66} & \multicolumn{1}{c|}{58.91} & 38.22 & \multicolumn{1}{c|}{79.61} & \multicolumn{1}{c|}{81.57} & \multicolumn{1}{c|}{72.63} & 25.40 & \multicolumn{1}{c|}{69.54} & \multicolumn{1}{c|}{79.00} & \multicolumn{1}{c|}{66.77} & 50.03 & \multicolumn{1}{c|}{86.23} & \multicolumn{1}{c|}{46.13} & \multicolumn{1}{c|}{31.00} & \textbf{54.27} & \multicolumn{1}{c|}{93.75} & \multicolumn{1}{c|}{77.73} & \multicolumn{1}{c|}{70.74} & 34.24 \\ \cline{2-22} 
\multicolumn{1}{|c|}{} & \textbf{AD} & \multicolumn{1}{c|}{86.54} & \multicolumn{1}{c|}{72.41} & \multicolumn{1}{c|}{61.58} & \textbf{38.94} & \multicolumn{1}{c|}{80.58} & \multicolumn{1}{c|}{75.88} & \multicolumn{1}{c|}{64.64} & 18.24 & \multicolumn{1}{c|}{75.63} & \multicolumn{1}{c|}{78.37} & \multicolumn{1}{c|}{64.35} & 54.77 & \multicolumn{1}{c|}{78.26} & \multicolumn{1}{c|}{45.21} & \multicolumn{1}{c|}{32.98} & 44.52 & \multicolumn{1}{c|}{79.17} & \multicolumn{1}{c|}{75.12} & \multicolumn{1}{c|}{60.20} & 33.70 \\ \cline{2-22} 
\multicolumn{1}{|c|}{} & \textbf{GH} & \multicolumn{1}{c|}{87.50} & \multicolumn{1}{c|}{67.31} & \multicolumn{1}{c|}{57.14} & 38.66 & \multicolumn{1}{c|}{78.64} & \multicolumn{1}{c|}{77.60} & \multicolumn{1}{c|}{69.13} & 17.46 & \multicolumn{1}{c|}{52.79} & \multicolumn{1}{c|}{75.42} & \multicolumn{1}{c|}{64.08} & 47.39 & \multicolumn{1}{c|}{92.75} & \multicolumn{1}{c|}{49.22} & \multicolumn{1}{c|}{35.16} & 53.70 & \multicolumn{1}{c|}{93.75} & \multicolumn{1}{c|}{84.62} & \multicolumn{1}{c|}{78.73} & 34.38 \\ \cline{2-22} 
\multicolumn{1}{|c|}{} & \textbf{SO} & \multicolumn{1}{c|}{89.90} & \multicolumn{1}{c|}{59.43} & \multicolumn{1}{c|}{51.15} & 38.82 & \multicolumn{1}{c|}{88.35} & \multicolumn{1}{c|}{83.29} & \multicolumn{1}{c|}{72.62} & 26.03 & \multicolumn{1}{c|}{85.28} & \multicolumn{1}{c|}{84.92} & \multicolumn{1}{c|}{70.87} & \textbf{61.33} & \multicolumn{1}{c|}{94.93} & \multicolumn{1}{c|}{52.92} & \multicolumn{1}{c|}{38.52} & 53.34 & \multicolumn{1}{c|}{89.58} & \multicolumn{1}{c|}{70.90} & \multicolumn{1}{c|}{59.79} & \textbf{35.77} \\ \hline
\end{tabular}}
\end{table*}

%% file: figure/win_counts.tex
\begin{figure*}[t!]
\pgfplotsset{compat=1.18}
\centering
\begin{tikzpicture}
\begin{groupplot}[
    group style={
        group name=my plots, 
        group size=4 by 1, 
        xlabels at=edge bottom, 
        ylabels at=edge left,
        vertical sep = 0.5cm
    },
    height=0.22\linewidth, 
    width=0.27\linewidth, 
    ylabel={Win Counts},
    ybar, 
    enlargelimits=0.15, 
    symbolic x coords={A, B, C, D, E, F}, 
    xtick=\empty,
]

\nextgroupplot[
    title={a. Basic RAG vs BI},
    bar width = 0.15cm,
    nodes near coords,
    nodes near coords style={
        font=\fontsize{5pt}{4pt}\selectfont 
    },
    legend style={
        at={(1.565, -0.36)}, 
        anchor=south, 
        legend columns=6, 
        /tikz/every even column/.append style={column sep=0.2cm},
        legend cell align={left}
    },
    ymin=2.6,
    ymax=20,
]
\addplot+[bar shift=0cm, fill=green!30, draw=green!80, text=green!80] coordinates {(A,13)}; \addlegendentry{Cmb vs BI}
\addplot+[bar shift=0cm, fill=teal!70, draw=teal!100, text=teal!100] coordinates {(B,12)}; \addlegendentry{API vs BI}
\addplot+[bar shift=0cm, fill=violet!30, draw=violet!80, text=violet!80] coordinates {(C,14)}; \addlegendentry{GH vs BI}
\addplot+[bar shift=0cm, fill=black!70, draw=black!100, text=black!100] coordinates {(D,11)}; \addlegendentry{SO vs BI}

\nextgroupplot[
    title={b. API RAG vs BI},
    bar width = 0.15cm,
    nodes near coords,
    nodes near coords style={
        font=\fontsize{5pt}{4pt}\selectfont 
    },
    ymin=2.6,
    ymax=20
]
\addplot+[bar shift=0cm, fill=green!30, draw=green!80, text=green!80] coordinates {(A,15)}; 
\addplot+[bar shift=0cm, fill=teal!70, draw=teal!100, text=teal!100] coordinates {(B,13)}; 
\addplot+[bar shift=0cm, fill=violet!30, draw=violet!80, text=violet!80] coordinates {(C,14)}; 
\addplot+[bar shift=0cm, fill=black!70, draw=black!100, text=black!100] coordinates {(D,12)};

\nextgroupplot[
    title={c. Basic RAG},
    bar width = 0.15cm,
    nodes near coords,
    nodes near coords style={
        font=\fontsize{5pt}{4pt}\selectfont 
    },
    legend style={
        at={(-0.28, -0.66)}, 
        anchor=south, 
        legend columns=6, 
        /tikz/every even column/.append style={column sep=0.2cm},
        legend cell align={left}
    },
    ymin=2.6,
    ymax=20
]
\addplot+[bar shift=0cm, fill=blue!30, draw=blue!80, text=blue!80] coordinates {(A,11)}; \addlegendentry{Cmb. vs API}
\addplot+[bar shift=0cm, fill=blue!70, draw=blue!100, text=blue!100] coordinates {(B,10)}; \addlegendentry{Cmb vs GH}
\addplot+[bar shift=0cm, fill=red!30, draw=red!80, text=red!80] coordinates {(C,12)}; \addlegendentry{Cmb vs SO}
\addplot+[bar shift=0cm, fill=red!70, draw=red!100, text=red!100] coordinates {(D,9)}; \addlegendentry{API vs GH}
\addplot+[bar shift=0cm, fill=brown!50, draw=brown!100, text=brown!100] coordinates {(E,9)}; \addlegendentry{API vs SO}
\addplot+[bar shift=0cm, fill=olive, draw=olive, text=olive] coordinates {(F,10)}; \addlegendentry{GH vs SO}

\nextgroupplot[
    title={d. API RAG},
    bar width = 0.15cm,
    nodes near coords,
    nodes near coords style={
        font=\fontsize{5pt}{4pt}\selectfont 
    },
    ymin=2.6,
    ymax=20
]
\addplot+[bar shift=0cm, fill=blue!30, draw=blue!80, text=blue!80] coordinates {(A,17)}; 
\addplot+[bar shift=0cm, fill=blue!70, draw=blue!100, text=blue!100] coordinates {(B,8)}; 
\addplot+[bar shift=0cm, fill=red!30, draw=red!80, text=red!80] coordinates {(C,13)}; 
\addplot+[bar shift=0cm, fill=red!70, draw=red!100, text=red!100] coordinates {(D,4)}; 
\addplot+[bar shift=0cm, fill=brown!50, draw=brown!100, text=brown!100] coordinates {(E,7)}; 
\addplot+[bar shift=0cm, fill=olive, draw=olive, text=olive] coordinates {(F,12)}; 
\end{groupplot}

\end{tikzpicture}
\caption{The left two sub-figures show the win counts of the RAG approaches vs the basic instruction prompting (BI). Cmb denotes combined RAG, API denotes API documents, GH denotes GitHub issues, and SO denotes StackOverflow Q\&As. The right two sub-figures show the win counts within the RAG approaches.}
\label{fig:win_counts}
\end{figure*}

%% file: table/api_a.tex
\begin{table*}[t!]
\caption{RQ3: Result of LLMs using API RAG prompting.
}
\vspace{-0.1in}
\label{tab:api_a}
\setlength{\tabcolsep}{3pt}
\resizebox{\textwidth}{!}{
\begin{tabular}{cc|cccc||cccc||cccc||cccc||cccc|}
\cline{3-22}
\multicolumn{1}{l}{} & \multicolumn{1}{l|}{} & \multicolumn{4}{c||}{\textbf{tf}} & \multicolumn{4}{c||}{\textbf{torch}} & \multicolumn{4}{c||}{\textbf{sklearn}} & \multicolumn{4}{c||}{\textbf{xgb}} & \multicolumn{4}{c|}{\textbf{jax}} \\ \hline
\multicolumn{1}{|c|}{\textbf{LLM}} & \textbf{BL} & \multicolumn{1}{c|}{\textbf{PR\%}} & \multicolumn{1}{c|}{\textbf{EX\%}} & \multicolumn{1}{c|}{\textbf{PS\%}} & \textbf{CV} & \multicolumn{1}{c|}{\textbf{PR\%}} & \multicolumn{1}{c|}{\textbf{EX\%}} & \multicolumn{1}{c|}{\textbf{PS\%}} & \textbf{CV} & \multicolumn{1}{c|}{\textbf{PR\%}} & \multicolumn{1}{c|}{\textbf{EX\%}} & \multicolumn{1}{c|}{\textbf{PS\%}} & \textbf{CV} & \multicolumn{1}{c|}{\textbf{PR\%}} & \multicolumn{1}{c|}{\textbf{EX\%}} & \multicolumn{1}{c|}{\textbf{PS\%}} & \textbf{CV} & \multicolumn{1}{c|}{\textbf{PR\%}} & \multicolumn{1}{c|}{\textbf{EX\%}} & \multicolumn{1}{c|}{\textbf{PS\%}} & \textbf{CV} \\ \hline
\multicolumn{1}{|c|}{\multirow{5}{*}{\textbf{4o}}} & \textbf{BI} & \multicolumn{1}{c|}{100.00} & \multicolumn{1}{c|}{89.17} & \multicolumn{1}{c|}{81.29} & 39.08 & \multicolumn{1}{c|}{100.00} & \multicolumn{1}{c|}{91.09} & \multicolumn{1}{c|}{84.49} & 25.66 & \multicolumn{1}{c|}{100.00} & \multicolumn{1}{c|}{92.51} & \multicolumn{1}{c|}{80.22} & 68.19 & \multicolumn{1}{c|}{100.00} & \multicolumn{1}{c|}{45.70} & \multicolumn{1}{c|}{33.33} & 56.93 & \multicolumn{1}{c|}{100.00} & \multicolumn{1}{c|}{89.38} & \multicolumn{1}{c|}{83.95} & 34.45 \\ \cline{2-22} 
\multicolumn{1}{|c|}{} & \textbf{CB} & \multicolumn{1}{c|}{100.00} & \multicolumn{1}{c|}{82.70} & \multicolumn{1}{c|}{72.99} & \textbf{40.09} & \multicolumn{1}{c|}{100.00} & \multicolumn{1}{c|}{90.08} & \multicolumn{1}{c|}{83.24} & 26.38 & \multicolumn{1}{c|}{100.00} & \multicolumn{1}{c|}{84.27} & \multicolumn{1}{c|}{69.17} & \textbf{69.54} & \multicolumn{1}{c|}{100.00} & \multicolumn{1}{c|}{72.77} & \multicolumn{1}{c|}{57.18} & \textbf{58.71} & \multicolumn{1}{c|}{95.83} & \multicolumn{1}{c|}{83.05} & \multicolumn{1}{c|}{75.71} & 35.88 \\ \cline{2-22} 
\multicolumn{1}{|c|}{} & \textbf{AD} & \multicolumn{1}{c|}{100.00} & \multicolumn{1}{c|}{84.96} & \multicolumn{1}{c|}{75.57} & 39.37 & \multicolumn{1}{c|}{100.00} & \multicolumn{1}{c|}{94.34} & \multicolumn{1}{c|}{94.34} & 19.18 & \multicolumn{1}{c|}{100.00} & \multicolumn{1}{c|}{82.57} & \multicolumn{1}{c|}{70.31} & 69.82 & \multicolumn{1}{c|}{100.00} & \multicolumn{1}{c|}{66.60} & \multicolumn{1}{c|}{49.90} & 57.65 & \multicolumn{1}{c|}{100.00} & \multicolumn{1}{c|}{80.92} & \multicolumn{1}{c|}{76.33} & 34.92 \\ \cline{2-22} 
\multicolumn{1}{|c|}{} & \textbf{GH} & \multicolumn{1}{c|}{100.00} & \multicolumn{1}{c|}{80.76} & \multicolumn{1}{c|}{70.06} & 40.28 & \multicolumn{1}{c|}{100.00} & \multicolumn{1}{c|}{91.62} & \multicolumn{1}{c|}{82.81} & \textbf{31.89} & \multicolumn{1}{c|}{100.00} & \multicolumn{1}{c|}{85.14} & \multicolumn{1}{c|}{72.83} & 49.59 & \multicolumn{1}{c|}{100.00} & \multicolumn{1}{c|}{59.07} & \multicolumn{1}{c|}{45.57} & 57.73 & \multicolumn{1}{c|}{100.00} & \multicolumn{1}{c|}{89.68} & \multicolumn{1}{c|}{80.52} & 35.33 \\ \cline{2-22} 
\multicolumn{1}{|c|}{} & \textbf{SO} & \multicolumn{1}{c|}{100.00} & \multicolumn{1}{c|}{77.37} & \multicolumn{1}{c|}{69.76} & 38.83 & \multicolumn{1}{c|}{100.00} & \multicolumn{1}{c|}{94.07} & \multicolumn{1}{c|}{89.17} & 25.76 & \multicolumn{1}{c|}{100.00} & \multicolumn{1}{c|}{91.67} & \multicolumn{1}{c|}{80.28} & 66.15 & \multicolumn{1}{c|}{100.00} & \multicolumn{1}{c|}{65.77} & \multicolumn{1}{c|}{51.34} & 55.19 & \multicolumn{1}{c|}{100.00} & \multicolumn{1}{c|}{89.18} & \multicolumn{1}{c|}{81.97} & \textbf{35.98} \\ \hline \hline
\multicolumn{1}{|c|}{\multirow{5}{*}{\textbf{3.5T}}} & \textbf{BI} & \multicolumn{1}{c|}{99.04} & \multicolumn{1}{c|}{71.84} & \multicolumn{1}{c|}{60.58} & 36.72 & \multicolumn{1}{c|}{93.20} & \multicolumn{1}{c|}{88.01} & \multicolumn{1}{c|}{76.40} & 25.43 & \multicolumn{1}{c|}{96.45} & \multicolumn{1}{c|}{91.41} & \multicolumn{1}{c|}{74.48} & 59.76 & \multicolumn{1}{c|}{99.28} & \multicolumn{1}{c|}{64.08} & \multicolumn{1}{c|}{49.51} & \textbf{48.12} & \multicolumn{1}{c|}{100.00} & \multicolumn{1}{c|}{81.75} & \multicolumn{1}{c|}{77.78} & 33.14 \\ \cline{2-22} 
\multicolumn{1}{|c|}{} & \textbf{CB} & \multicolumn{1}{c|}{96.15} & \multicolumn{1}{c|}{58.77} & \multicolumn{1}{c|}{49.13} & \textbf{38.91} & \multicolumn{1}{c|}{93.20} & \multicolumn{1}{c|}{82.67} & \multicolumn{1}{c|}{67.17} & 31.27 & \multicolumn{1}{c|}{96.45} & \multicolumn{1}{c|}{71.06} & \multicolumn{1}{c|}{50.92} & \textbf{22.28} & \multicolumn{1}{c|}{93.48} & \multicolumn{1}{c|}{64.21} & \multicolumn{1}{c|}{51.58} & 36.23 & \multicolumn{1}{c|}{91.67} & \multicolumn{1}{c|}{62.14} & \multicolumn{1}{c|}{55.71} & \textbf{34.52} \\ \cline{2-22} 
\multicolumn{1}{|c|}{} & \textbf{AD} & \multicolumn{1}{c|}{96.63} & \multicolumn{1}{c|}{72.38} & \multicolumn{1}{c|}{63.35} & 37.47 & \multicolumn{1}{c|}{94.17} & \multicolumn{1}{c|}{89.10} & \multicolumn{1}{c|}{78.19} & 25.50 & \multicolumn{1}{c|}{96.45} & \multicolumn{1}{c|}{73.91} & \multicolumn{1}{c|}{62.88} & 21.67 & \multicolumn{1}{c|}{99.28} & \multicolumn{1}{c|}{40.41} & \multicolumn{1}{c|}{30.05} & 44.11 & \multicolumn{1}{c|}{97.92} & \multicolumn{1}{c|}{66.08} & \multicolumn{1}{c|}{62.57} & 33.41 \\ \cline{2-22} 
\multicolumn{1}{|c|}{} & \textbf{GH} & \multicolumn{1}{c|}{96.15} & \multicolumn{1}{c|}{54.43} & \multicolumn{1}{c|}{42.20} & 38.53 & \multicolumn{1}{c|}{98.06} & \multicolumn{1}{c|}{80.25} & \multicolumn{1}{c|}{64.20} & \textbf{33.21} & \multicolumn{1}{c|}{95.94} & \multicolumn{1}{c|}{61.29} & \multicolumn{1}{c|}{38.71} & 22.11 & \multicolumn{1}{c|}{100.00} & \multicolumn{1}{c|}{51.11} & \multicolumn{1}{c|}{44.44} & 43.59 & \multicolumn{1}{c|}{95.83} & \multicolumn{1}{c|}{59.85} & \multicolumn{1}{c|}{44.70} & 34.59 \\ \cline{2-22} 
\multicolumn{1}{|c|}{} & \textbf{SO} & \multicolumn{1}{c|}{91.35} & \multicolumn{1}{c|}{47.11} & \multicolumn{1}{c|}{38.45} & 37.22 & \multicolumn{1}{c|}{97.09} & \multicolumn{1}{c|}{81.49} & \multicolumn{1}{c|}{61.21} & 25.90 & \multicolumn{1}{c|}{96.45} & \multicolumn{1}{c|}{96.45} & \multicolumn{1}{c|}{96.45} & 96.45 & \multicolumn{1}{c|}{97.83} & \multicolumn{1}{c|}{26.23} & \multicolumn{1}{c|}{20.49} & 32.88 & \multicolumn{1}{c|}{77.08} & \multicolumn{1}{c|}{77.08} & \multicolumn{1}{c|}{41.49} & \textbf{36.61} \\ \hline \hline
\multicolumn{1}{|c|}{\multirow{5}{*}{\textbf{MS}}} & \textbf{BI} & \multicolumn{1}{c|}{95.19} & \multicolumn{1}{c|}{42.73} & \multicolumn{1}{c|}{32.04} & 37.05 & \multicolumn{1}{c|}{95.15} & \multicolumn{1}{c|}{64.19} & \multicolumn{1}{c|}{50.68} & 18.29 & \multicolumn{1}{c|}{92.39} & \multicolumn{1}{c|}{69.67} & \multicolumn{1}{c|}{54.38} & \textbf{58.49} & \multicolumn{1}{c|}{95.65} & \multicolumn{1}{c|}{38.46} & \multicolumn{1}{c|}{25.34} & \textbf{51.52} & \multicolumn{1}{c|}{97.92} & \multicolumn{1}{c|}{75.98} & \multicolumn{1}{c|}{61.57} & 33.01 \\ \cline{2-22} 
\multicolumn{1}{|c|}{} & \textbf{CB} & \multicolumn{1}{c|}{82.21} & \multicolumn{1}{c|}{45.65} & \multicolumn{1}{c|}{28.23} & 37.16 & \multicolumn{1}{c|}{78.64} & \multicolumn{1}{c|}{70.20} & \multicolumn{1}{c|}{48.48} & 30.79 & \multicolumn{1}{c|}{73.60} & \multicolumn{1}{c|}{67.42} & \multicolumn{1}{c|}{50.47} & 55.88 & \multicolumn{1}{c|}{89.86} & \multicolumn{1}{c|}{44.53} & \multicolumn{1}{c|}{33.59} & 47.57 & \multicolumn{1}{c|}{79.17} & \multicolumn{1}{c|}{70.51} & \multicolumn{1}{c|}{58.97} & \textbf{34.30} \\ \cline{2-22} 
\multicolumn{1}{|c|}{} & \textbf{AD} & \multicolumn{1}{c|}{83.65} & \multicolumn{1}{c|}{42.97} & \multicolumn{1}{c|}{31.18} & 36.97 & \multicolumn{1}{c|}{95.15} & \multicolumn{1}{c|}{67.65} & \multicolumn{1}{c|}{56.30} & 25.78 & \multicolumn{1}{c|}{81.73} & \multicolumn{1}{c|}{71.08} & \multicolumn{1}{c|}{57.70} & 55.57 & \multicolumn{1}{c|}{86.96} & \multicolumn{1}{c|}{38.05} & \multicolumn{1}{c|}{32.74} & 44.81 & \multicolumn{1}{c|}{81.25} & \multicolumn{1}{c|}{59.79} & \multicolumn{1}{c|}{47.42} & 33.54 \\ \cline{2-22} 
\multicolumn{1}{|c|}{} & \textbf{GH} & \multicolumn{1}{c|}{81.73} & \multicolumn{1}{c|}{41.18} & \multicolumn{1}{c|}{27.78} & \textbf{37.78} & \multicolumn{1}{c|}{83.50} & \multicolumn{1}{c|}{59.50} & \multicolumn{1}{c|}{43.50} & \textbf{32.35} & \multicolumn{1}{c|}{74.62} & \multicolumn{1}{c|}{55.09} & \multicolumn{1}{c|}{37.89} & 51.79 & \multicolumn{1}{c|}{88.41} & \multicolumn{1}{c|}{35.19} & \multicolumn{1}{c|}{16.05} & 47.09 & \multicolumn{1}{c|}{79.17} & \multicolumn{1}{c|}{65.28} & \multicolumn{1}{c|}{41.67} & 35.12 \\ \cline{2-22} 
\multicolumn{1}{|c|}{} & \textbf{SO} & \multicolumn{1}{c|}{75.00} & \multicolumn{1}{c|}{37.98} & \multicolumn{1}{c|}{26.87} & 36.50 & \multicolumn{1}{c|}{73.79} & \multicolumn{1}{c|}{69.27} & \multicolumn{1}{c|}{48.78} & 25.50 & \multicolumn{1}{c|}{64.47} & \multicolumn{1}{c|}{69.54} & \multicolumn{1}{c|}{53.00} & 57.41 & \multicolumn{1}{c|}{75.36} & \multicolumn{1}{c|}{26.24} & \multicolumn{1}{c|}{18.55} & 45.74 & \multicolumn{1}{c|}{66.67} & \multicolumn{1}{c|}{66.10} & \multicolumn{1}{c|}{44.07} & \textbf{36.19} \\ \hline \hline
\multicolumn{1}{|c|}{\multirow{5}{*}{\textbf{LM}}} & \textbf{BI} & \multicolumn{1}{c|}{80.77} & \multicolumn{1}{c|}{50.28} & \multicolumn{1}{c|}{43.50} & 32.84 & \multicolumn{1}{c|}{85.44} & \multicolumn{1}{c|}{79.09} & \multicolumn{1}{c|}{71.52} & 17.43 & \multicolumn{1}{c|}{64.97} & \multicolumn{1}{c|}{80.00} & \multicolumn{1}{c|}{70.00} & \textbf{22.78} & \multicolumn{1}{c|}{94.93} & \multicolumn{1}{c|}{55.09} & \multicolumn{1}{c|}{55.56} & 16.72 & \multicolumn{1}{c|}{87.50} & \multicolumn{1}{c|}{66.67} & \multicolumn{1}{c|}{55.56} & 30.95 \\ \cline{2-22} 
\multicolumn{1}{|c|}{} & \textbf{CB} & \multicolumn{1}{c|}{85.58} & \multicolumn{1}{c|}{59.50} & \multicolumn{1}{c|}{61.21} & \textbf{39.31} & \multicolumn{1}{c|}{93.20} & \multicolumn{1}{c|}{89.27} & \multicolumn{1}{c|}{81.57} & 32.16 & \multicolumn{1}{c|}{71.57} & \multicolumn{1}{c|}{79.55} & \multicolumn{1}{c|}{66.56} & 18.73 & \multicolumn{1}{c|}{93.48} & \multicolumn{1}{c|}{64.21} & \multicolumn{1}{c|}{51.58} & 36.23 & \multicolumn{1}{c|}{87.50} & \multicolumn{1}{c|}{73.83} & \multicolumn{1}{c|}{65.89} & 34.21 \\ \cline{2-22} 
\multicolumn{1}{|c|}{} & \textbf{AD} & \multicolumn{1}{c|}{69.71} & \multicolumn{1}{c|}{72.21} & \multicolumn{1}{c|}{61.18} & 38.10 & \multicolumn{1}{c|}{90.29} & \multicolumn{1}{c|}{89.33} & \multicolumn{1}{c|}{81.17} & 18.96 & \multicolumn{1}{c|}{72.08} & \multicolumn{1}{c|}{69.08} & \multicolumn{1}{c|}{60.40} & 19.85 & \multicolumn{1}{c|}{72.46} & \multicolumn{1}{c|}{20.59} & \multicolumn{1}{c|}{17.65} & 29.97 & \multicolumn{1}{c|}{66.67} & \multicolumn{1}{c|}{70.69} & \multicolumn{1}{c|}{66.09} & 33.29 \\ \cline{2-22} 
\multicolumn{1}{|c|}{} & \textbf{GH} & \multicolumn{1}{c|}{79.81} & \multicolumn{1}{c|}{65.13} & \multicolumn{1}{c|}{55.00} & 38.94 & \multicolumn{1}{c|}{88.35} & \multicolumn{1}{c|}{84.81} & \multicolumn{1}{c|}{71.03} & \textbf{32.90} & \multicolumn{1}{c|}{61.93} & \multicolumn{1}{c|}{76.67} & \multicolumn{1}{c|}{58.89} & 19.32 & \multicolumn{1}{c|}{84.78} & \multicolumn{1}{c|}{44.83} & \multicolumn{1}{c|}{32.76} & \textbf{41.65} & \multicolumn{1}{c|}{89.58} & \multicolumn{1}{c|}{80.79} & \multicolumn{1}{c|}{70.94} & 34.84 \\ \cline{2-22} 
\multicolumn{1}{|c|}{} & \textbf{SO} & \multicolumn{1}{c|}{90.87} & \multicolumn{1}{c|}{59.03} & \multicolumn{1}{c|}{53.08} & 38.93 & \multicolumn{1}{c|}{95.15} & \multicolumn{1}{c|}{89.58} & \multicolumn{1}{c|}{81.66} & 26.01 & \multicolumn{1}{c|}{85.79} & \multicolumn{1}{c|}{67.30} & \multicolumn{1}{c|}{52.13} & 20.82 & \multicolumn{1}{c|}{86.23} & \multicolumn{1}{c|}{81.82} & \multicolumn{1}{c|}{65.15} & 35.66 & \multicolumn{1}{c|}{85.42} & \multicolumn{1}{c|}{77.51} & \multicolumn{1}{c|}{65.09} & \textbf{37.26} \\ \hline
\end{tabular}}
\end{table*}

%% file: table/BD_dist.tex
\begin{table}[t!]
\centering
\caption{\respto{3-1a} \respto{5-6b} \add{Results of bug detection. TC denotes test case (method-level). EX denotes Execution Rate, PS denotes Pass Rate, and BD denotes Bug Detectability. Basic refers to basic instruction prompting. A-RAG refers to API-level API documentation RAG.}} 
\vspace{-0.1in}
\label{tab:bd_dist}
\resizebox{1.03\linewidth}{!}{
\begin{tabular}{|c|cc||cc||cc||cc||cc|}
\hline
\textbf{Library} & \multicolumn{2}{c||}{\textbf{tf}} & \multicolumn{2}{c||}{\textbf{torch}} & \multicolumn{2}{c||}{\textbf{sklearn}} & \multicolumn{2}{c||}{\textbf{jax}} & \multicolumn{2}{c|}{\textbf{xgb}} \\ \hline
\textbf{Prompting} & \multicolumn{1}{c|}{\textbf{Basic}} & \textbf{A-RAG} & \multicolumn{1}{c|}{\textbf{Basic}} & \textbf{A-RAG} & \multicolumn{1}{c|}{\textbf{Basic}} & \textbf{A-RAG} & \multicolumn{1}{c|}{\textbf{Basic}} & \textbf{A-RAG} & \multicolumn{1}{c|}{\textbf{Basic}} & \textbf{A-RAG} \\ \hline
\textbf{Total APIs} & \multicolumn{2}{c||}{208} & \multicolumn{2}{c||}{103} & \multicolumn{2}{c||}{197} & \multicolumn{2}{c||}{48} & \multicolumn{2}{c|}{138} \\ \hline
\textbf{Total TCs} & \multicolumn{1}{c|}{1,355} & 1,239 & \multicolumn{1}{c|}{603} & 530 & \multicolumn{1}{c|}{1,213} & 1,028 & \multicolumn{1}{c|}{255} & 240 & \multicolumn{1}{c|}{324} & 402 \\ \hline
\textbf{Failed TCs} & \multicolumn{1}{c|}{167} & 168 & \multicolumn{1}{c|}{77} & 50 & \multicolumn{1}{c|}{205} & 192 & \multicolumn{1}{c|}{15} & 21 & \multicolumn{1}{c|}{95} & 95 \\ \hline
\textbf{EX (\%)} & \multicolumn{1}{c|}{83.99} & 83.21 & \multicolumn{1}{c|}{87.40} & 87.17 & \multicolumn{1}{c|}{89.78} & 88.91 & \multicolumn{1}{c|}{87.84} & 87.50 & \multicolumn{1}{c|}{78.40} & 72.39 \\ \hline
\textbf{PS (\%)} & \multicolumn{1}{c|}{71.66} & 69.65 & \multicolumn{1}{c|}{74.63} & 77.74 & \multicolumn{1}{c|}{72.88} & 70.23 & \multicolumn{1}{c|}{81.96} & 78.75 & \multicolumn{1}{c|}{49.07} & 48.76 \\ \hline
\textbf{BD} & \multicolumn{1}{c|}{1} & 10 & \multicolumn{1}{c|}{4} & 3 & \multicolumn{1}{c|}{4} & 5 & \multicolumn{1}{c|}{0} & 0 & \multicolumn{1}{c|}{0} & 1 \\ \hline
\end{tabular}
}
\end{table}

%% file: figure/bug_ex.tex
\lstset{ 
  basicstyle=\ttfamily\tiny,        
  breaklines=true,                 
  commentstyle=\color{brown},    
  firstnumber=1,                
  frame=single,	                   
  keepspaces=true,                 
  keywordstyle=\color{blue},       
  numbers=left,                    
  numbersep=7pt,                   
  numberstyle=\tiny\color{orange}, 
  rulecolor=\color{black},         
  showspaces=false,                
  showstringspaces=false,          
  showtabs=false,                  
  stepnumber=1,                    
  stringstyle=\color{purple},     
  tabsize=2,	                   
  title=\lstname                   
}
\begin{figure}[t!]
\centering
\begin{lstlisting}[language=python]
# Generated unit test for tf.boolean_mask
...
def test_mask_with_different_dtype(self):
    # Test with a mask of different dtype (int)
    tensor = tf.constant([1, 2, 3, 4, 5])
    mask = tf.constant([1, 0, 1, 0, 1], dtype=tf.int32)
    with self.assertRaises(TypeError):
        tf.boolean_mask(tensor, mask)
...
# Error Log
======================================================================
FAIL: test_mask_with_different_dtype (__main__.TestBooleanMask)
----------------------------------------------------------------------
Traceback (most recent call last):
  File "/tf.boolean_mask.py", line 28, in test_mask_with_different_dtype
    tf.boolean_mask(tensor, mask)
AssertionError: TypeError not raised
\end{lstlisting}
\vspace{-0.1in}
\caption{An example bug detected by a generated test.}
\label{lst:bug_ex}
\end{figure}

%% file: figure/venn.tex
\begin{figure}[t!]
    \centering
    \includegraphics[width=0.8\linewidth]{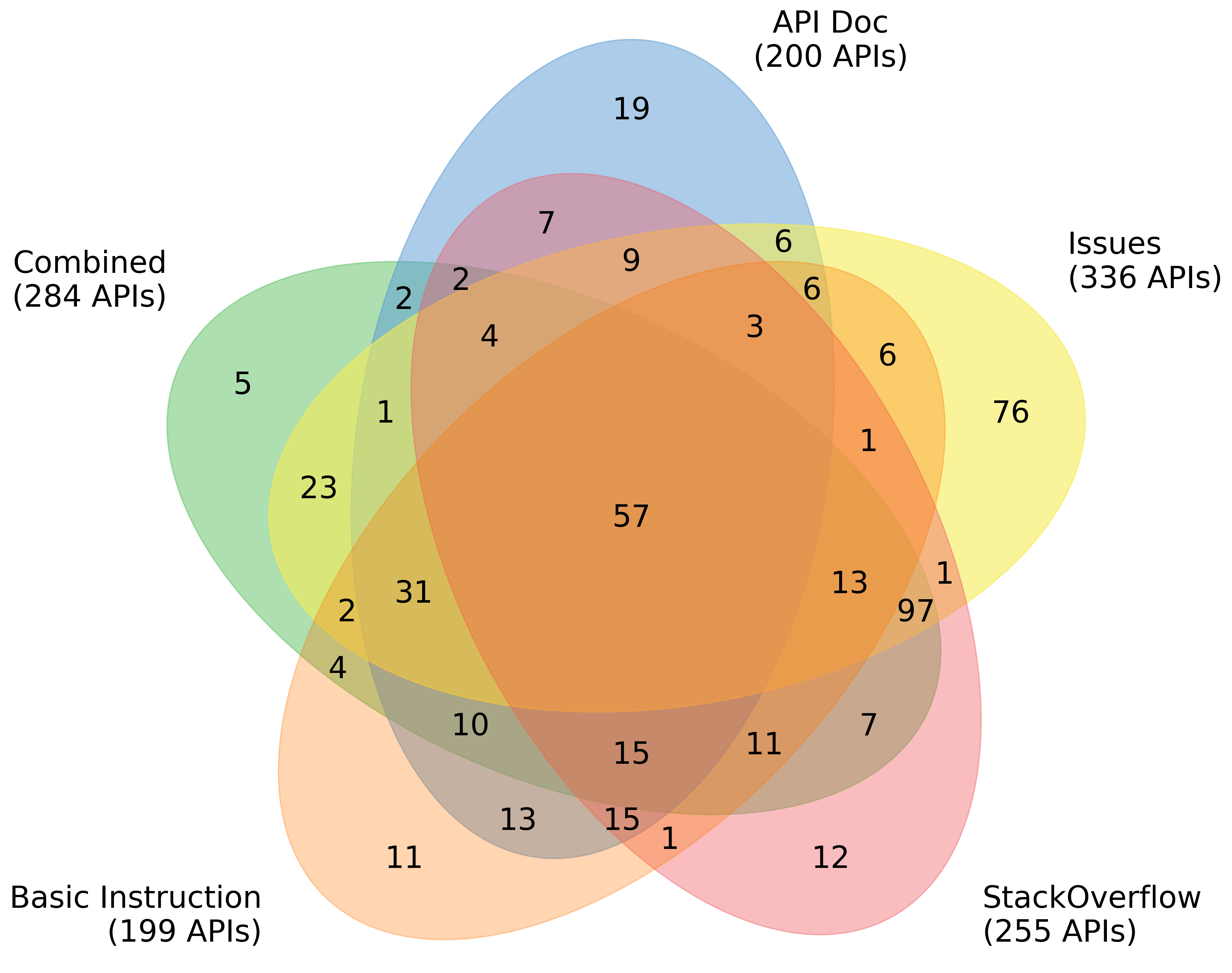}
    \vspace{-0.2in}
    \caption{The number of APIs with unique coverage.}
    \label{fig:venn}
\end{figure}

%% file: figure/bar.tex
\begin{figure}[t!]
    \centering
    \includegraphics[width=0.8\linewidth]{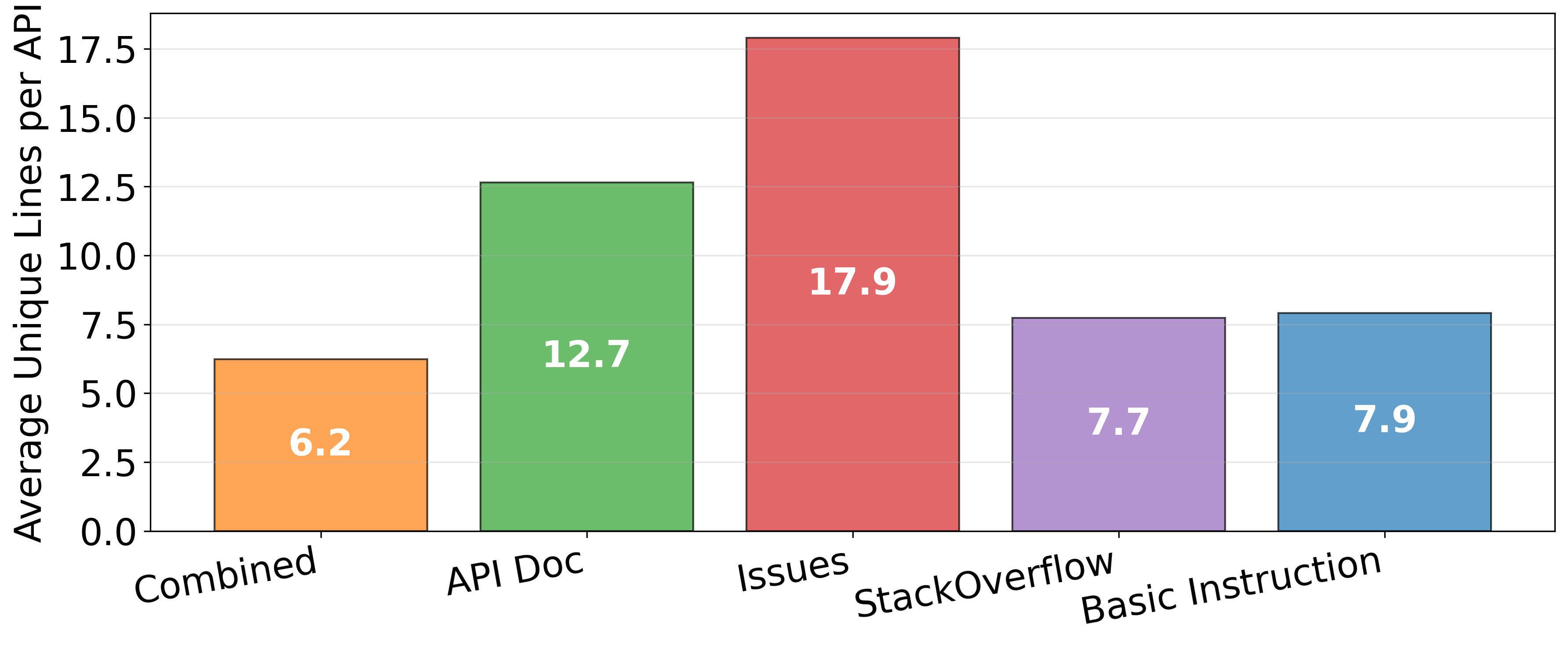}
    \vspace{-0.2in}
    \caption{The average number of unique lines covered per API by each approach.}
    \label{fig:bar}
\end{figure}

%% file: table/manual_category.tex
\begin{table}[t!]
\centering
\caption{\respto{4-1} \add{RQ5: The number of documents in each information  category. Bold denotes the highest count for a category compared to other RAGs.}
}
\vspace{-0.15in}
\label{tab:category}
\resizebox{\linewidth}{!}{
\begin{tabular}{|r|c|c|c|c||c|}
\hline
\textbf{Categories} & \textbf{API Docs} & \textbf{GH Issues} & \textbf{SO Q\&As} & \textbf{Combined} & \textbf{Total} \\ \hline
\textbf{API sequence} & 35 (19\%) & 86 (21\%) & \textbf{99} (33\%) & 79 (23\%) & 299 \\ \hline
\textbf{Bug-inducing} & - & \textbf{72} (18\%) & 6 (2\%) & 25 (7\%) & 103 \\ \hline
\textbf{Exception} & 37 (20) & \textbf{49} (12\%) & 27 (9\%) & 40 (12\%) & 153 \\ \hline
\textbf{Integration} & 1 (1\%) & 26 (6\%) & \textbf{33} (11\%) & 20 (6\%) & 80 \\ \hline
\textbf{Performance} & 1 (1\%) & \textbf{12} (3\%) & 6 (2\%) & 2 (1\%) & 21 \\ \hline
\textbf{Standard cases} & 27 (15\%) & 78 (19\%) & 78 (26\%) & \textbf{96} (28\%) & 279 \\ \hline
\textbf{Unique input} & 80 (44\%) & \textbf{82} (20\%) & 55 (18\%) & 81 (24\%) & 298 \\ \hline \hline
\textbf{Total} & 181 & 405 & 304 & 343 & 1233 \\ \hline
\end{tabular}
}
\end{table}

%% file: sections/5.threats.tex
\section{Threats to Validity}
\label{sec:threats}
\noindent \textbf{Internal Validity.} 
The primary threat to internal validity arises from the nondeterministic nature of LLMs. 
To mitigate this, we set the temperature parameter to zero, thereby constraining randomness and ensuring the most deterministic response possible \cite{ouyang2023llm}. Additionally, we conducted a preliminary study by running a subset of the experiment five times (using only the TensorFlow and PyTorch libraries) and found that the average difference in metrics was less than 2.5\%.  
There may also be limitations in the prompts used for the LLMs. To address these, we conducted multiple series of prompt optimizations before applying them to the entire dataset, ensuring the prompts were as effective as possible.

\noindent \textbf{Construct Validity.} 
The primary threat to construct validity lies in the evaluation metrics employed. The test adequacy metric used in this study is line coverage. Although line coverage is widely utilized to assess the effectiveness of test cases, it may not strongly correlate with the detection of actual bugs, which is the ultimate goal of testing. The mutation score is another widely used metric in the domain, designed to detect mutations (potential defects) in the software under test. However, in our case, mutating and building the source projects is too time-consuming, rendering our study infeasible. Therefore, we decided to add a manual analysis of the bug detectability of the unit test cases to mitigate this. Further investigation on using fault-based testing in this matter necessitates additional future studies. \respto{1-3} \add{Another threat is that we did not explicitly evaluate the relevance of the retrieved documents for the different RAGs. In the interest of maintaining scope, we evaluated the end-to-end test coverage and bug-finding performance, rather than examining detailed retrieval accuracy metrics. Consequently, we will address this aspect in future work.} \respto{8-8} \add{Lastly, our benchmarks rely on Python‐level coverage tools and thus do not exercise core logic implemented in C/C++ extensions. Instrumenting these lower‐level components would require specialized tooling and deep integration with their build systems, which is beyond our current scope. Consequently, our coverage metrics reflect only the front‐end API surface; comprehensive C/C++ level testing remains important for future work.}

\noindent \textbf{Conclusion Validity.}
We conducted both a quantitative and qualitative study to investigate the effect of different RAG approaches and the detection of bugs in the generated unit test cases. However, we examined a subset of libraries and APIs, which could pose a potential threat to the validity of our claims.
\respto{8-6} \add{A further threat arises from the reliance on closed-source LLM APIs (e.g., \emph{GPT-3.5-Turbo, GPT-4o}) that regularly update without public visibility or transparency. Although we cannot control these models, the model version used for our experiment is preserved to replicate as much as possible.}

\noindent \textbf{External Validity.} 
The main threats to external validity in this study include limitations related to (a) LLMs, (b) the projects, (c) the dataset, and (d) the programming language investigated. 
Regarding LLMs, we utilized the most widely used state-of-the-art models in various sizes, considering our budget constraints. Including other LLM baselines might alter the observations from this study and will be worth exploring when they become available in the future. 
We selected five projects for this study, but these do not encompass all projects available.
Concerning the dataset, we tried to mine all relevant information from the API documentation, GitHub, and StackOverflow. Nevertheless, there may be missing information that we failed to mine. We also limited the documents to those relevant to the APIs using string-matching heuristics. 
To make the experiment manageable in terms of budget and execution time, we used APIs that have at least 10 documents for GitHub issues and StackOverflow Q\&As, as sufficient documents are needed for RAG to function. This approach excluded a significant amount of data from the experiment; however, with more resources and data, this could be further improved in future studies. Despite these limitations, we have devised sufficient projects, instances, and documents to demonstrate significant observations from the study. 

%% file: sections/6.conclusion.tex
\section{Conclusion}
\label{sec:conclusion}

This paper evaluates different RAG approaches for generating unit tests in ML/DL libraries, comparing Basic and API-level RAG using API docs, GitHub issues, and Stack Overflow. Our results indicate that GitHub issues are particularly effective, often outperforming other sources and basic instruction prompting in terms of code coverage. Combining RAG sources further improves results. Our manual analysis shows LLM-generated tests can find real bugs. While LLMs handle general cases without extra docs, adding unique external examples helps cover untested code lines. Future work includes improving retrieval algorithms for corner cases and exploring broader knowledge sources like specification documents and technical blogs to enhance automated testing. 
